\documentclass[printer]{aa}

\bibpunct{(}{)}{;}{a}{}{,}

\usepackage{graphicx}
\usepackage{txfonts}
\usepackage{soul}
\usepackage{txfonts}
\usepackage{color}
\usepackage{ulem}

\newcommand{\hii}{H\textsc{ii}}
\def\ks{km s$^{-1}$}

\def\s{$^{\prime\prime}$}

\def\cm3{cm$^{-3}$}

\def\2{$^{12}$CO}
\def\3{$^{13}$CO}
\def\8{C$^{18}$O}
\def\msol{M$_\odot$}

\def\cm2{cm$^{-2}$}

\begin{document}

\title{Looking for evidence of high-mass star formation at core scale in a  massive molecular clump}

\author {M. E. Ortega \inst{1}
\and N. C. Martinez \inst{1}
\and S. Paron \inst{1}
\and A. Marinelli \inst{1}
\and N. L. Isequilla \inst{1}
}

\institute{CONICET - Universidad de Buenos Aires. Instituto de Astronom\'{\i}a y F\'{\i}sica del Espacio
             CC 67, Suc. 28, 1428 Buenos Aires, Argentina\\
             \email{mortega@iafe.uba.ar}
}

\offprints{M. E. Ortega}

   \date{Received <date>; Accepted <date>}

\abstract{High-mass stars are formed as a result of the fragmentation of massive molecular clumps. However, what it is not clear is whether this fragmentation gives rise to stable prestellar cores massive enough to form directly high-mass stars or leads to prestellar cores of low masses that, by acquiring material from the environment, generate high-mass stars.  Nowadays, several observational studies focus on the characterization of prestellar massive clumps candidates. Nevertheless, studies of active massive clumps at different evolutionary stages are still needed to gain a complete understanding of how high-mass stars form.}{We present a comprehensive physical and chemical study of the fragmentation and star formation activity towards the massive  clump AGAL G338.9188+0.5494 harbouring the extended green object EGO 338.92+0.55(b). The presence of an EGO embedded in a massive clump, suggests, at clump scale, that high-mass star formation is occurring. The main goal of this work is to find evidence of such high-mass star formation, but at core scale.}
{Using millimeter observations of continuum and molecular lines obtained from the Atacama Large Millimeter Array database at Bands 6 and 7, we study the substructure of the massive clump AGAL G338.9188+0.5494. The angular resolution of the data at Band 7 is about 0\farcs5, which allow us to resolve structures of about 0.01~pc ($\sim$ 2000~au) at the distance of 4.4~kpc.} 
{The continuum emission at 340~GHz reveals that the molecular clump is fragmented in five cores, labeled from C1 to C5. The $^{12}$CO J=3--2 emission shows the presence of molecular outflows related to three of them. The molecular outflow related to the core C1 is among the most massive (from 0.25 to 0.77~\msol) and energetic (from $0.4 \times 10^{46}$ to $1.2 \times 10^{46}$~erg), considering studies carried out with similar observations towards this type of sources.
Rotational diagrams for the CH$_3$CN and CH$_3$CCH yield temperatures of about 340 and 72~K, respectively, for the core C1. The different temperatures show that the methyl cyanide would trace a gas layer closer to the protostar than the methyl acetylene, which would trace outermost layers.

Using a range of temperatures going from 120 K (about the typical molecular desorption temperature in hot cores) to the temperature derived from CH$_3$CN (about 340~K), the mass of core C1 ranges  from 3 to 10 \msol. 
The mid-IR 4.5~$\mu$m extended emission related to the EGO coincides in position and inclination with the discovered molecular outflow arising from core C1, which indicate that it should be 
the main responsible for the 4.5 $\mu$m brightness. The average mass and energy of such a molecular outflow is about 0.5~\msol~and $10^{46}$~erg, respectively, which suggest that 10 \msol~is the most likely mass value for core C1. Additionally we found that the region is chemically very rich with several complex molecular species. Particularly, from the analysis of the CN emission we found strong evidence that such a radical is indirectly tracing the molecular outflows,more precisely the border of the cavity walls carved out by such outflows,
and hence we point out that this is probably the first clear detection of CN
as a tracer of molecular outflows in star-forming regions. }{}

\titlerunning{.}
\authorrunning{M. E. Ortega et al.}


\maketitle
%

\section{Introduction}

The formation of a high-mass star begins with the fragmentation of a massive clump into smaller structures known as molecular cores. However, what is not clear, is whether this fragmentation gives rise to prestellar cores massive enough (a few tens of solar masses) to form directly these stars or leads to cores of low and intermediate masses that generate high-mass stars, acquiring material from their environment \citep{palau2018, moscadelli2021}. In the first scenario, high-mass stars form through an individual monolithic core collapse, in the second one, they form from a global hierarchical collapse of a massive clump, where many low and intermediate mass cores competitively accrete material from the surrounding through converging gas filaments that feed the cores \citep{motte18, sch19}. These outlines of the high-mass star formation scenarios overlook several aspects of their chemical and physical complexity, which are treated in detail in the following reviews: \citet{krumholz2009}, \citet{tan2014}, and \citet{vazquez2019}.

Nowadays, an important research line in the field of high-mass star formation is focused on studying the fragmentation of
massive clumps in their earliest stages. The main goal is to detect the presence of massive pre-stellar cores, however, it is still a matter of debate if they exist and, in such a case, if they are stable enough against further fragmentation to give rise to the formation of a high-mass star through monolithic collapse.

 \begin{figure*}[tt]
   \centering
   \includegraphics[width=18cm]{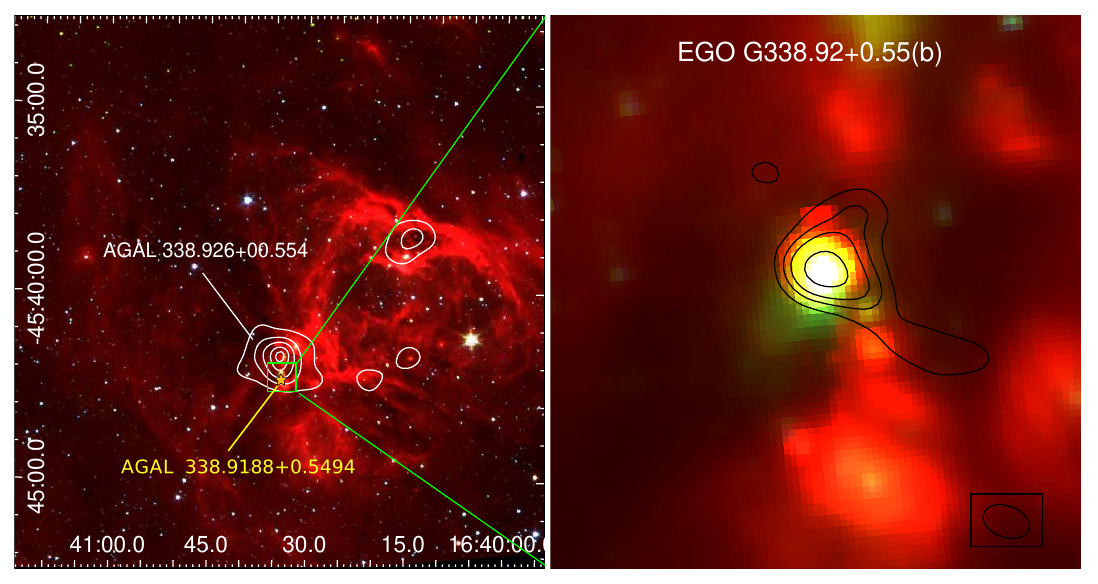}
    \caption{Large scale surrounding of EGO G338.92+0.55(b). Left-panel: Overview of the \hii~region G338.90+00.60 at Spitzer 3.6 (blue), 4.5 (green), 8.0 (red)~$\mu$m bands. The white contours represent the continuum emission at 870~$\mu$m extracted from the ATLASGAL survey. Levels are: 1, 3, 6, 9, and 12~Jy beam$^{-1}$. The green square highlights the  region studied in this work, which includes the position of EGO G338.92+0.55(b) represented by the yellow star. Right-panel: A close-up view of the EGO at the same Spitzer bands. The black contours represent the ALMA continuum emission at 340~GHz (7~m array). Levels are: 0.2, 0.4, 0.7, and 1.4 Jy beam$^{-1}$. The beam of the ALMA observation is indicated at the bottom-right corner. The blue, green, and red colour scales go from 10 to 500~MJy sr$^{-1}$.}
    \label{intro}
\end{figure*}

Several studies based on data from Atacama Large Millimeter Array (ALMA) of infrared-quiet massive clumps have revealed different fragmentation properties: in some cases, limited fragmentation (very few cores and with super-Jeans masses, well above the solar mass) with a large fraction of pre-stellar cores with masses in the range 8--120~\msol~\citep[e.g.;][]{wang14, csengeri2017, neupane2020}, and, in other cases, a large population of low--mass ($\leq$~1~\msol) pre-stellar cores with a maximum core mass of 11~\msol~\citep{sanhueza2019}.  In this regard, \citet{kai13} pointed out that a possible explanation for the different fragmentation characteristics could be the size-scale-dependent collapse time-scale that results from the finite size of real molecular clouds, which is indeed predicted by analytical models \citep{pon11}.

\begin{table*}[tt]
\small
\centering
\caption{Main ALMA data parameters of the bands 6 and 7 in the 12~m array.}
\label{data_info}
\begin{tabular}{lcccccccc}
\hline\hline
  Project                & ALMA band & Freq. range & Beam size & Line sens. (10~\ks) & $\Delta\nu$ & $\Delta$v & Max. reco. scale & FOV\\
                         & & (GHz) & (\s $\times$ \s) & (mJy beam$^{-1}$) & (MHz) & (\ks) & (arcsec) & (arcsec) \\
\hline                 
2015.1.01312 & 6 & 224.2 $\sim$ 242.7 & 0.72 $\times$ 0.68 & 1.5 & 1.1 & 1.4 & 6.1 & 25\\
2017.1.00914 & 7 & 333.4 $\sim$ 349.1 & 0.48 $\times$ 0.46 & 3.6$^{\bf *}$ & 1.1 & 1.0 & 5.1 & 32\\
\hline
\multicolumn{8}{l}{\tiny {\bf *} The continuum sensitivity at 340~GHz is about 0.2~mJy beam$^{-1}$.}\\
\end{tabular}
\end{table*}

\citet{csengeri2017} carried out a fragmentation study towards a sample of ATLASGAL sources identified as infrared-quiet massive clumps using continuum ALMA data at a spatial resolution of about 0.06~pc. The authors found limited fragmentation towards most of the sources. According to the authors, a possible explanation could be that early fragmentation of massive clumps does not follow thermal processes, which leads to fragment masses largely exceeding the local Jeans mass. Thus, a combination of turbulence, magnetic field, and radiative feedback would be increasing the necessary mass for fragmentation.  Another explanation could be that these early stages could correspond to a phase of compactness where the large level of fragmentation to form a cluster has not yet developed.

Among the sources characterized by \citet{csengeri2017} is AGAL G035.1330$-$00.745 (hereafter AGAL35), towards which the authors identified two molecular cores. Assuming an average clump temperature of 25~K, they derived masses of about 36 and 8~\msol.   

However,  \citet{ortega2022}, based on continuum and line ALMA data with a spatial resolution of about 0.007~pc, identified four molecular cores towards AGAL35. Besides, the authors estimated masses below 2~\msol~for the cores, using core temperatures above 100~K derived from the CH$_3$CN J=13--12 transition. The authors also found molecular outflow activity towards two molecular cores.  They concluded that considering an average clump temperature for the estimation of the masses of the cores could be inadequate even in the case of infrared-quiet massive clumps. This  assumption would be resulting in an overestimation of the masses of the cores. This study confirms that  a prestellar clump candidate can present star formation activity, manifested as hot cores and/or molecular outflows, when it is studied at the core scale. Paraphrasing \citet{pillai2019}, we wonder whether  high extinction can hide very young and low-luminosity protostars within such seemingly starless clumps, and thus, are the existing cases of high-mass starless  cores starless?
  
Finding massive prestellar cores stable against further fragmentation would support the monolithic collapse scenario, however it is not an easy task. On the other hand,  it is equally important to carry out detailed characterizations of massive clumps with recent star formation activity like the presence of hot molecular cores (HMCs) to achieve a more complete picture of how the fragmentation occurs. Additionally, HMCs are the chemically richest regions in the interstellar medium (ISM) (e.g. \citealt{bonfand19,herbst09}), and the star forming processes strongly influence the chemistry of such environments \citep{jorgen20}. Hence, observing molecular lines and studying their emission and chemistry is important to  characterize physical and chemical conditions of the gas and, eventually, to figure out the evolutive stage of the fragmentation. 

Nowadays there are not many works in the literature that connect high-mass star formation signatures at the clump scale with evidence of high-mass star formation at the core scale. A good candidate to carry out such kind of study would be a massive molecular clump harbouring an Extended Green Object (EGO). \citet{cyga2008} catalogued more than 300 EGOs, based on their extended 4.5~$\mu$m emission in GLIMPSE images. EGOs are defined as  massive young stellar objects (MYSOs) candidates to harbour molecular o0utflows. Thus, the presence of a bright EGO embedded in a massive clump suggests that high-mass star formation is taking place. Keeping this in mind, we searched for ALMA observations\footnote{https://almascience.nrao.edu} towards the brightest EGOs in the catalogue with the additional requirement that these EGOs are associated with a high-mass ATLASGAL source. Furthermore, it is essential for this study that the ALMA data include molecules from which accurate temperature values can be obtained for the cores. Thus, the EGO 338.92$+$0.55(b) (hereafter EGO\,G338) embedded in the massive clump AGAL G338.9188$+$0.5494 was the selected source, which is presented in detail in the next section.

\section{Presentation of the region}
\label{present}

The submillimeter source AGAL G338.926$+$00.554 \citep{contreras2013} is located towards the eastern border of the \hii~region G338.90$+$00.60 (see Figure\,\ref{intro} left-panel). 
\citet{wienen2015}, based on $^{13}$CO J=1--0 emission, estimated a systemic velocity of about $-$64.1~\ks~for this ATLASGAL source, which correspond to a near kinematic distance of about 4.4~kpc.

\citet{csengeri2014} extended the ATLASGAL \citet{contreras2013}'s catalog of compact sources  using an optimized source extraction method. The authors identified a total of 10861 compact submillimeter clumps, increasing by far the number of previously detected sources. Particularly, we figured out that AGAL G338.926$+$00.554 is composed by four minor dust condensations. Among them, AGAL G338.9188$+$0.5494 (hereafter AGAL 338; see yellow star in Fig. \ref{intro}-left panel), is the dust condensation associated with EGO\,G338. 

Figure\,\ref{intro}-right panel shows a close–up view of the location of EGO\,G338 at the same mid-infrared bands. The black contours represent the ALMA submillimeter continuum emission at 340 GHz (in the 7~m array) with an angular resolution of about 4$\arcsec$~(see Sect.\,\ref{ALMAdata}. It can be appreciated a conspicuous dust condensation in positional coincidence with the peak of emission at the mid-infrared bands associated with EGO\,G338.

\section{Data}
\label{ALMAdata}

The data cubes from the projects 2015.1.01312 (PI: Fuller, G.; Band 6) and 2017.1.00914 (PI: Csengeri, T.; Band 7) were obtained from the ALMA Science Archive\footnote{http://almascience.eso.org/aq/}.
The single pointing observations for the target were carried out using the following telescope configurations with L5BL/L80BL(m): 42.6/221.3 for project 2015.1.01312 and 34.5/226.8 for project 2017.1.00914, in the 12~m array in both cases. Table \ref{data_info} shows the main ALMA data parameters.

Project 2017.1.00914 also includes observations of Band 7 in the 7~m array with angular resolution and continuum sensitivity of about 3$\farcs$7 and 1.2~mJy beam$^{-1}$, respectively, and a telescope configuration with L5BL/L80BL(m):8.7/27.5. We only used the continuum at 340~GHz in the 7~m array to introduce the region (see Fig.\,\ref{intro}-right panel). Then, along the manuscript, when we refer to 340 GHz continuum, we are referring to the 12m-array.

We extracted all the molecular lines from the Band 6, except the $^{12}$CO J=3--2 transition, which, together with the continuum at 340~GHz, was obtained from Band 7.

It is important to remark that even though the data of both projects passed the QA2 quality level, which assures a reliable calibration for a ``science ready'' data, the automatic pipeline imaging process may give raise to a clean image with some artefacts. For example, an inappropriate setting of the parameters of the {\it clean} task in CASA could generate artificial dips in the spectra. Thus, we reprocessed the raw data using CASA 4.5.1 and 4.7.2 versions and the calibration pipelines scripts. Particular care was taken with the different parameters of the task {\it clean}. The images and spectra obtained from our data reprocessing, after several runs of the {\it clean} task varying some of its parameters, were very similar to those obtained from the archival. 

The task {\it imcontsub} in CASA was used to subtract the continuum from the spectral lines using a first order polynomial. The frequency ranges without molecular line emission were carefully selected in each spectral window. The continuum map at 340~GHz in the 12~m array was obtained averaging the continuum emission of each of the four spectral windows and was corrected for primary beam. Several continuum subtraction tests, which involved the selection of different free line regions of the spectra, were performed to ensure a reliable 340~GHz continuum map. This map has an rms noise level of about 0.2 mJy beam$^{-1}$.

Given that high-spatial resolution is required to properly characterize the clump fragmentation and star formation activity at core scales, it is important to remark that the beam size of the 340~GHz continuum data in the 12~m array provides a spatial resolution of about 0.01~pc ($\sim$ 2000 au) at the distance of 4.4~kpc, which is appropriate to spatially resolve the substructure of the clump AGAL 338.

\section{Results}
\label{results}

In the following subsections we present studies of fragmentation and star formation at core scales towards AGAL\,338 using the ALMA data at bands 6 and 7 in the 12~m array.

\subsection{Continuum emission: tracing the fragmentation}  
\label{dust}

\begin{figure}[h]
   \centering
   \includegraphics[width=9.2cm]{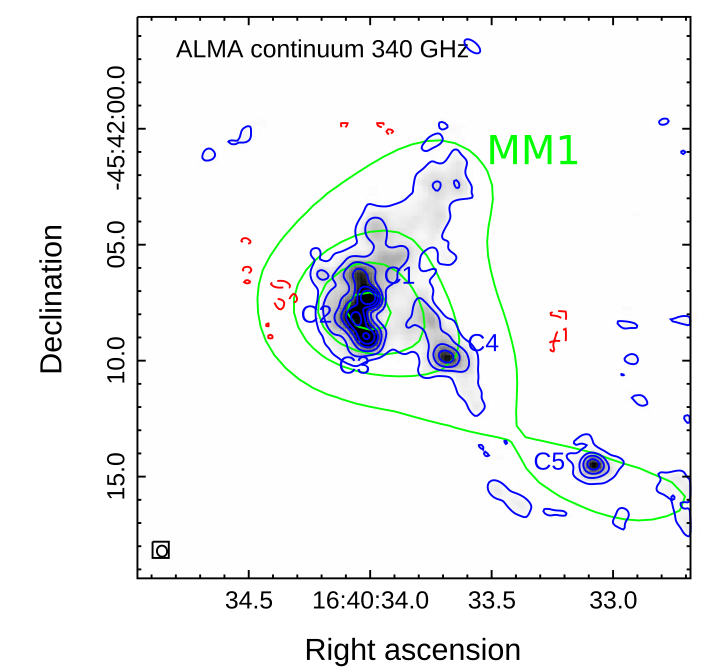}
    \caption{ALMA continuum emission at 340~GHz (12~m array). The grayscale goes from 1 to 180~mJy~beam$^{-1}$. The blue contours levels are: 1 (about 5~$\sigma$), 10, 30, 60, 90, and 140~mJy beam$^{-1}$. 
     The dashed red contours correspond to $-$1 mJy~beam$^{-1}$. The beam of 340~GHz continuum emission (12~m array) is indicated at the bottom left corner. The green contours correspond to the ALMA continuum emission at 340 GHz (7~m array) shown in Fig.\,\ref{intro}, with levels at 0.2, 0.4, 0.7, and 1.4 Jy~beam$^{-1}$.}
    \label{cont-340}
\end{figure}

We begin the study of the fragmentation of the dust clump AGAL 338 analysing the high resolution and sensitivity submillimeter continuum emission map at Band 7 (array 12~m). Figure\,\ref{cont-340} shows the ALMA continuum emission at 340~GHz in grayscale and blue contours. The green contours represent the ALMA continuum emission at 340~GHz (7~m array) presented in Fig.\,\ref{intro}, in which it can be noticed the presence of the conspicuous dust condensation, labeled MM1 in Fig.\,\ref{cont-340}, a faint tail-like feature towards the southwest direction, and a lobe-like feature aligned, but opposite, with the extended emission at 4.5~$\mu$m in the southeast-northwest direction.

The better angular resolution of the 12~m array observations allows us to identify five dust cores towards AGAL\,338, which are labeled from C1 to C5. In particular, we can resolve the MM1 condensation in four cores (C1 to C4), while the core C5 lies onto the faint tail-like structure as seen in the continuum emission of the 7~m array. Also it can be noticed faint emission in positional coincidence with the lobe-like structure extending towards the northwest.

\begin{table}[h]
\caption{Dust cores parameters from the continuum emission at 340~GHz, using the 2D Gaussian fitting tool from CASA.}
\label{contparams}
\tiny
\centering
\begin{tabular}{cccccc}
\hline\hline
Core & RA & Dec. & $\Theta_{\rm size}$  &  ${\rm I_{peak}}$ & S\\
 & (J2000) & (J2000) & (arcsec)  & (mJy beam$^{-1}$) & (mJy)\\
\hline
C1 & 16:40:34.0 & $-$45:42:07.3 & 0.83$\times$0.81 & 269$\pm$25 & 580$\pm$61\\ 
C2 & 16:40:34.1 & $-$45:42:08.2 & 0.73$\times$0.64 & 155$\pm$15 & 406$\pm$38\\ 
C3 & 16:40:34.0 & $-$45:42:09.0 & 0.64$\times$0.62 & 150$\pm$17 & 309$\pm$34\\ 
C4 & 16:40:33.7 & $-$45:42:09.8 & 0.77$\times$0.54 &  66$\pm$7  & 131$\pm$14\\ 
C5 & 16:40:33.1 & $-$45:42:14.5 & 0.61$\times$0.52 &  87$\pm$8  & 126$\pm$12\\
\hline
\end{tabular}
\end{table}

Table\,\ref{contparams} presents the main parameters of the dust continuum cores observed at 340~GHz. Columns\,2 and\,3 give the absolute position, Col.\,4 the angular size, Cols.\,5, and\,6 show the peak intensity I${\rm _{peak}}$ and the integrated intensity S, respectively. 

The core sizes are at least a factor six smaller than the maximum recoverable scales of the observations, which ensures that all of the flux of the cores is recovered.

\subsection{$^{12}$CO and C$^{17}$O: tracing the outflow activity and the ambient gas}
\label{12CO}

We searched for molecular outflow activity related to the clump AGAL\,338 and, in particular, associated with the EGO\,338, using the $^{12}$CO emission. After a carefully inspection of the channels of the spectral window containing the $^{12}$CO J=3--2 transition, we found several extended structures that suggest the presence of molecular outflows related to some of the dust cores. 

\begin{figure}[h]
   \centering
   \includegraphics[width=9.5cm]{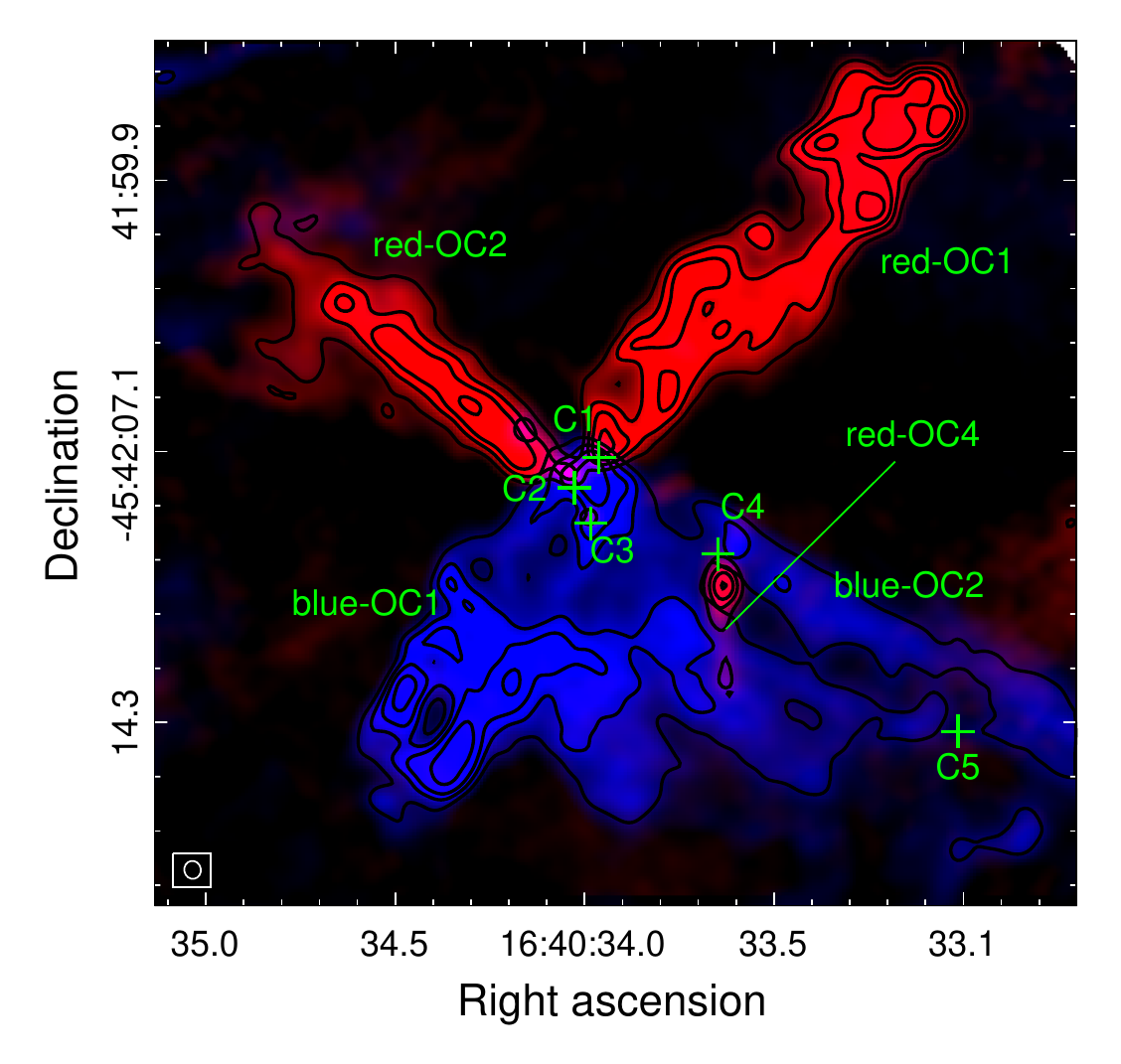}
    \caption{$^{12}$CO J=3--2 emission distribution integrated between $-$90 and $-$70 \ks~(blue), and between $-$55 and $+$5 \ks~(red). The systemic velocity of the complex is about $-$64 \ks. The contours levels are at 5, 8, 12, and 18~mJy beam$^{-1}$.
    The beam of the 340~GHz continuum and the $^{12}$CO J=3--2 emissions is the same and it is indicated at the bottom left corner.}
    \label{co-outflows}
\end{figure}

Figure\,\ref{co-outflows} shows the $^{12}$CO J=3--2 emission distribution integrated between $-$90 and $-$70 \ks~(blue), and between $-$55 and $+$5 \ks~(red). The systemic velocity of the complex is about $-$64~\ks~\citep{wienen2015}.

\begin{figure}[h]
   \centering
   \includegraphics[width=8.5cm]{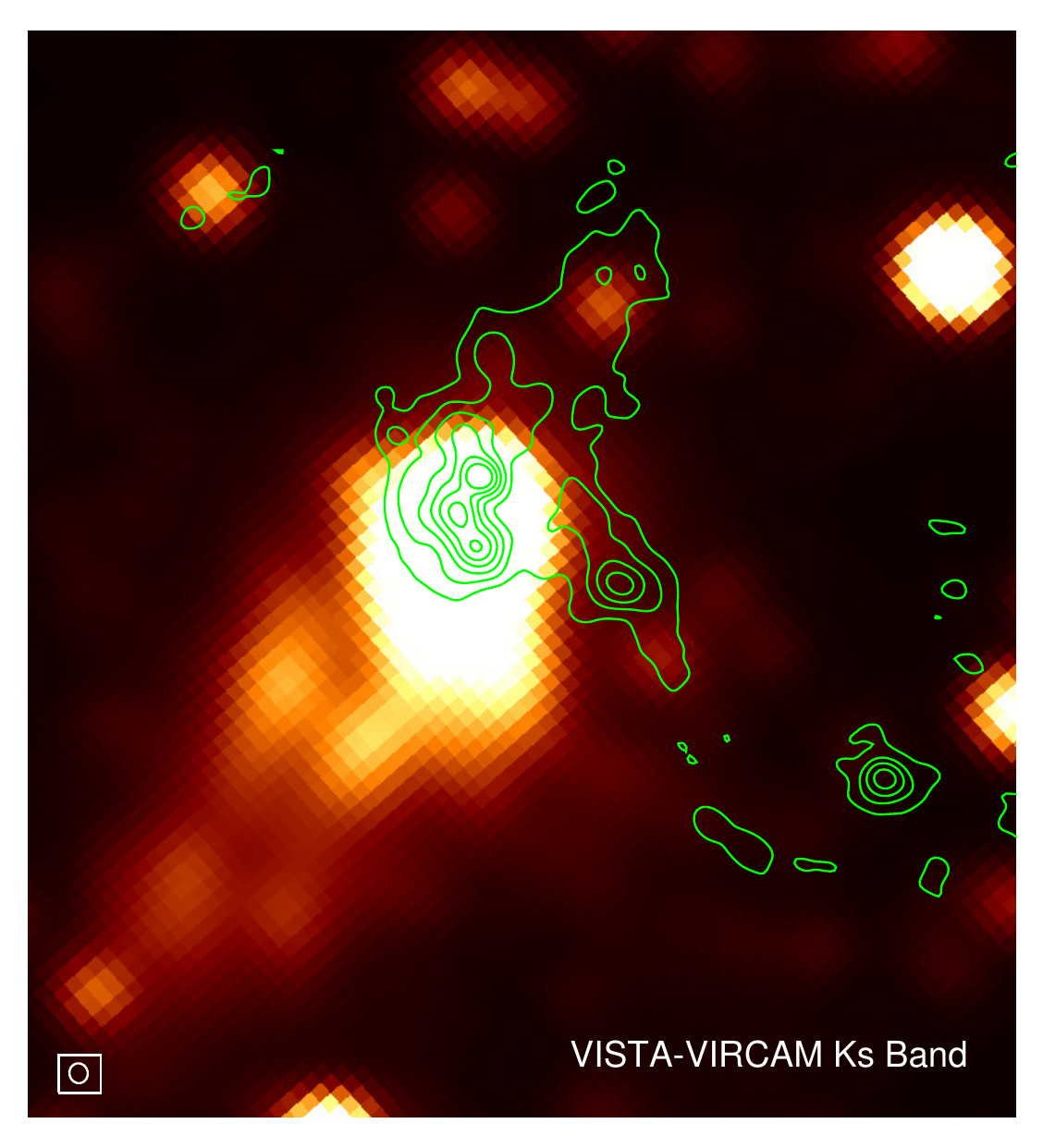}
    \caption{K${\rm _s}$ band emission from VISTA Hemisphere Survey with VIRCAM. The green contours represent the continuum emission at 340~GHz (12~m array). Levels are at 1, 10, 30, 60, 90, and 140~mJy beam$^{-1}$. The beam of the continuum emission  at 340~GHz is indicated at the bottom left corner.}
    \label{VISTA_Ks}
\end{figure}

\begin{figure}[h]
   \centering
   \includegraphics[width=8.3cm]{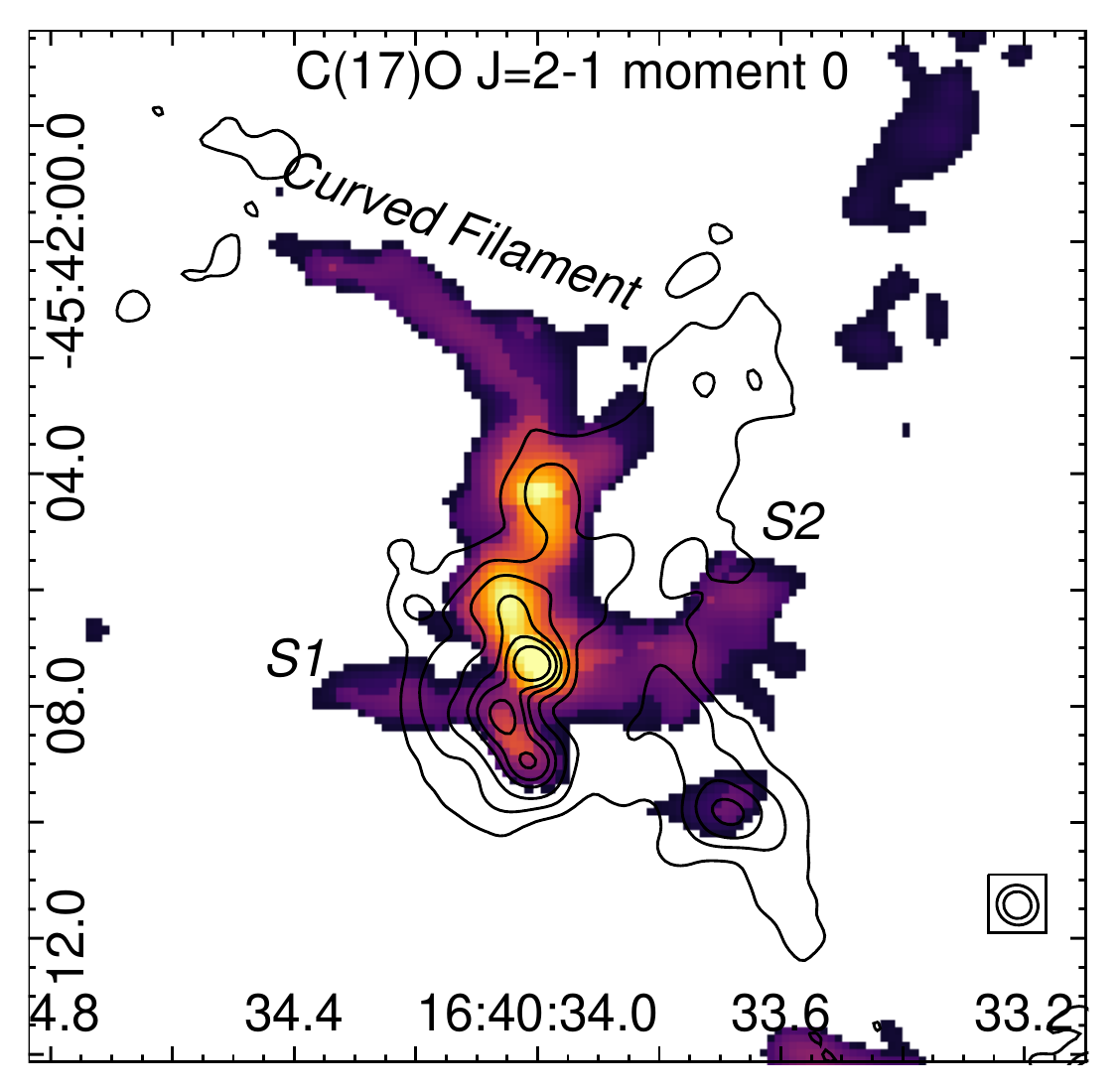}
   \includegraphics[width=8.5cm]{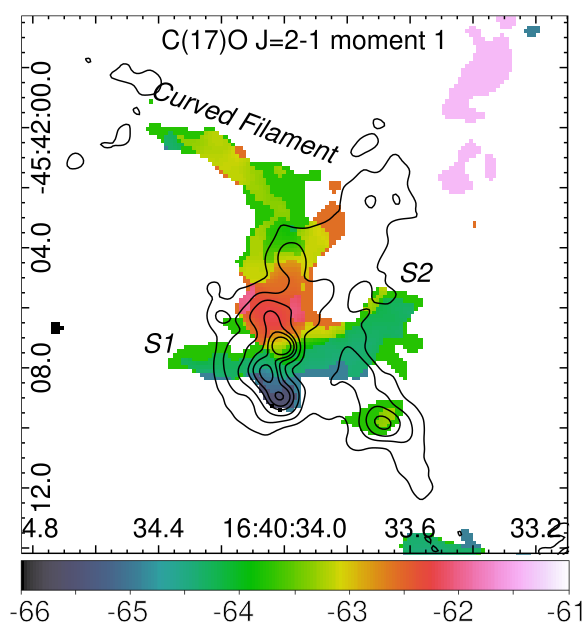}
    \caption{C$^{17}$O J=2--1 moment maps. Top panel: C$^{17}$O J=2--1 moment 0 map integrated between $-$66 and $-$61 \ks. Color-scale goes from 0.01 to 0.50~Jy beam$^{-1}$. The black contours represent the continuum emission at 340~GHz with levels at 1, 10, 30, 60, 90, and 140~mJy beam$^{-1}$. The beams of C$^{17}$O J=2--1 line and 340~GHz continuum emission are indicated at the bottom right corner. Bottom panel: C$^{17}$O J=2--1 moment 1 map. The units of the color-bar are \ks. The black contours represent the radio continuum at 340~GHz with levels at 1, 10, 30, 60, 90, and 140~mJy beam$^{-1}$. The systemic velocity is about $-64$~\ks. }
    \label{c17o}
\end{figure}

At first glance, it can be noticed an intense molecular outflow activity arising from the central region of the core cluster. The brightest core, C1, exhibits the most conspicuous molecular outflow, which is oriented in the southeast-northwest direction. The position of red-OC1 outflow coincides with the lobe-like structure extending towards the northwest mentioned in Sect.\,\ref{dust}, while the blue-OC1 outflow, less collimated, spatially coincides with the 4.5~$\mu$m extended emission of EGO\,G338 (see Fig. \ref{intro}-right-panel). The core C2 shows a more collimated outflow red-OC2, which can be appreciated in the northeast-southwest direction. From the contours displayed in Fig.\,\ref{co-outflows} it can be appreciated that the red-OC1 outflow is more clumpy than the red-OC2, which is the most collimated outflow in the region.
While both red outflows (red-OC1 and red-OC2) exhibit a clear spatial separation, the blue ones, blue-OC1 and blue-OC2, are blended and appear as a single cone-like shape structure that opens towards the south. 

Towards core C4, a faint red outflow appears extending southwards, likely associated with a weak and/or incipient outflow activity, while its blue counterpart seems to be contaminated by the blue-OC2 outflow.

A complete characterization of the  cores embedded in AGAL\,338, must include the study of the associated molecular outflow activity. Thus, following \citet{li2018}, we estimate the main parameters of red-OC1 and red-OC2 outflows. The blue counterparts, blue-OC1 and blue-OC2, are part of a single structure.

We calculate the column density and the total mass for each red lobe, using the following equations \citep{buckle10}: 

\begin{equation}
    {\rm N(^{12}CO)= 7.96 \times10^{13} \left(\frac{T_{ex}+0.92}{1-{exp{\left(\frac{-16.6}{T_{ex}}\right)}}}\right){exp\left(\frac{16.6}{T_{ex}}\right)} \int {\tau_{32} \mathrm{d}v}}
\end{equation}

\noindent and assuming that the $^{12}$CO J=3--2 emission is optically thin towards the outflows (e.g. \citealt{lebron06,shimo15}), 

\begin{equation}
   {\rm \int {\tau_{32} \mathrm{d}v} = {\frac{1}{J(T_{ex})-J(2.7~K)}} \int {T_{mb} \mathrm{d}\nu}}
\end{equation}

\begin{equation}
     {\rm with \hspace{0.4cm} J(T)= {\left(\frac{h\nu/k}{exp({\frac{h\nu}{kT}})-1}\right)}}
\end{equation}

\begin{equation}
    {\rm M_{out}= N(^{12}CO)~[H_{2}/CO]~\mu_{H_2}~m_{H}~A_{pixel}~N_{pixel}}
\end{equation}

\noindent where N($^{12}$CO) is the average column density of each lobe, ${\rm [CO/H_{2}]=10^{-4}}$ is the abundance ratio between the molecules, ${\rm \mu_{H_2}=2.72}$ is the mean molecular weight, ${\rm m_{H}=1.67 \times 10^{-24}}$~g is the mass of the hydrogen atom, ${\rm A_{pixel}}$ is the pixel area, and ${\rm N_{pixel}}$ is the pixels number that fills each lobe. The outflows parameters were estimated using  typical excitation temperatures ranging from 10 to 50~K \citep[e.g.;][]{li2020}. 

Table \ref{outflow_parameters} shows the main parameters derived for outflows red-OC1 and red-OC2: mass, momentum (${\rm P = M \bar v}$), energy, outflow mechanical force (F$_{\rm out}$=P/t$_{\rm dyn}$), length and dynamical age (${\rm t_{dyn} = Length/v_{max}}$), where ${\rm \bar v}$ and ${\rm v_{max}}$ are the median and maximum velocity of each velocity interval with respect to the systemic velocity of the gas associated with AGAL\,338.

\begin{table*}[tt]
\centering
\caption{Main parameters of the red-OC1 and red-OC2 molecular outflows for excitation temperatures of 10 and 50~K.}
\label{outflow_parameters}
\begin{tabular}{lcccc}
\hline\hline
Parameter  & \multicolumn{2}{c}{Red-OC1} & \multicolumn{2}{c}{Red-OC2} \\
                        \cline{2-3}\cline{4-5}
                        &  10 K  & 50 K & 10 K & 50 K \\
                        
\hline
Mass ( $\times 10^{-2}$ ${\rm M_{\odot}}$) & 77.2$\pm$26.3 & 25.3$\pm$7.1 & 28.5$\pm$8.6 & 8.1$\pm$3.2\\
Momentum (${\rm M_{\odot}~km~s^{-1}}$) & 30.8$\pm$12.1 & 10.0$\pm$4.2 & 11.1$\pm$4.3 & 3.5$\pm$1.2 \\
Energy (10$^{45}$ erg) & 12.2$\pm$6.3 & 3.9$\pm$2.1 & 4.4$\pm$2.5 & 1.4$\pm$0.6\\ 
F$_{\rm out}$ ($\times 10^{-3}$ ${\rm M_{\odot}~km~s^{-1}yr^{-1}}$) & 7.3$\pm$3.4 & 2.4$\pm$1.1 & 4.1$\pm$1.9 & 1.2$\pm$0.5 \\ 
\hline
Length (pc)   & \multicolumn{2}{c}{0.29$\pm$0.03} & \multicolumn{2}{c}{ 0.18$\pm$0.02} \\
Dynamical age ($\times 10^3$ yrs)  & \multicolumn{2}{c}{4.2$\pm$1.3} & \multicolumn{2}{c}{2.7$\pm$0.8}\\

\hline
\end{tabular}
\end{table*}

Additionally, we found that AGAL 338 shows near-IR emission at K${\rm _s}$ band associated with the EGO\,G338. Figure\,\ref{VISTA_Ks} presents the K${\rm _s}$-band emission obtained from the VISTA Hemisphere Survey \citep[VHS; ][]{mcmahon2013}. The green contours represent the continuum emission at 340~GHz. The bulk of emission coincides with the location of cores C1, C2, and C3. Conspicuous near-IR emission extends towards the southeast direction, which overlaps with the position of the outflow  blue-OC1 (see Fig.\,\ref{co-outflows}). However, there is no evidence of extended emission at near-IR related to the red-OC1 outflow.

Figure\,\ref{c17o}-top panel shows the C$^{17}$O J=2--1 moment 0 map integrated between $-$66 and $-$61 \ks. The contours represent the continuum emission at 340~GHz. The bulk of emission coincides with the location of core C1, {and is extended northwards, where another C$^{17}$O condensation appears in coincidence with a protrusion of the continuum emission (probably an incipient, or another not resolved core). Then a curved filament (indicated in Fig.\,\ref{c17o}) emerges further northwards. Additionally, it can be noticed two elongated structures, labeled S1 and S2, connected with the bulk of the emission, which extend from east to west.

Figure \ref{c17o}-bottom panel shows the C$^{17}$O J=2--1 moment 1 map integrated in the same velocity interval than the moment 0 map. The  C$^{17}$O J=2--1 associated with the core C1 is at the systemic velocity, while the nearby gas towards the north and the south shows red- and blue-shifted velocities, respectively, which could be tracing the birth of the molecular outflows. 
The curved filament does not show any velocity gradient, hence we discard that it can be a typical converging filament feeding the cores \citep{sch19,pineda22}.

\subsection{Analysis of molecular species} 
\label{molecular}

We analyze several molecular lines useful to characterize the physical conditions and the chemistry of the hot cores. The selected molecular lines are presented in Table\,\ref{molecLines}. Figure\,\ref{molecFig} presents the moment 0 maps of these molecular lines, showing the spatial distribution of the emission of each molecular species towards the region of EGO\,G338 in comparison with the dust millimeter continuum emission (displayed in green contours).

\begin{table}
\caption{Molecules and transitions shown in Fig. \ref{molecFig}.}
\label{molecLines}
\centering
\begin{tabular}{llcc}
\hline\hline
Molecule & Transition & Rest Freq. & E$_u$ \\
         &            &  (GHz)     & (K) \\
\hline
CH$_{3}$CN  & 13(2)--12(2)     &  239.119 & 108.9 \\
CH$_{3}$CCH & 14(2)--13(2) &        239.234 & 115.1 \\
HC$_{3}$N   & J=25--24  &           227.418 & 141.8 \\
H$_{2}$CS   & 7(3,5)--6(3,4)$^{\rm a}$ &  240.392 & 164.5 \\
CH$_{3}$OH  & 5(1,5)--4(1,4)$^{\rm a}$  &   239.746 &  49.1 \\
HNCO & 11(0,11)--10(0,10)$^{\rm b,c}$&  241.774 & 69.6 \\
CN   & N=2--1 J=5/2--3/2 $^{\rm b}$ &    &  \\
            & \hspace{0.3cm}(F=5/2-3/2) &  226.876  &  16.3     \\
C$^{34}$S   & J=5--4  &             241.016 & 27.7 \\
HDO        &  3(1,2)--2(2,1)  &     225.896 & 167.2 \\
\hline
\multicolumn{4}{l}{\tiny {a} Partially blended with another transition of the same molecule.}\\
\multicolumn{4}{l}{\tiny {b} Several hyperfine transitions blended.}\\
\multicolumn{4}{l}{\tiny {c} Partially blended with CH$_3$OH molecule.}\\
\end{tabular}
\end{table}

\begin{figure*}[h]
\includegraphics[width=6.2cm]{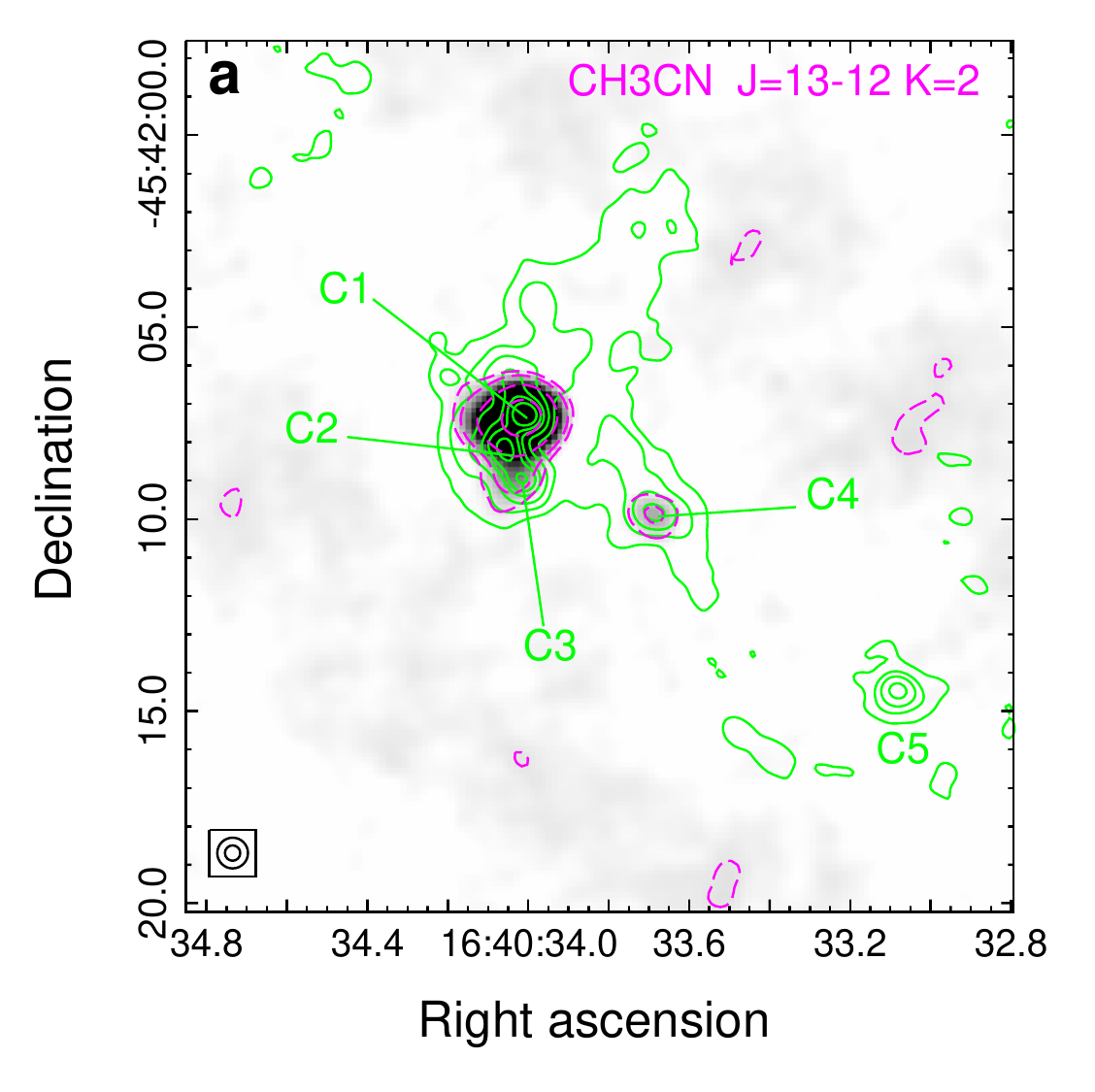}
\includegraphics[width=6.3cm]{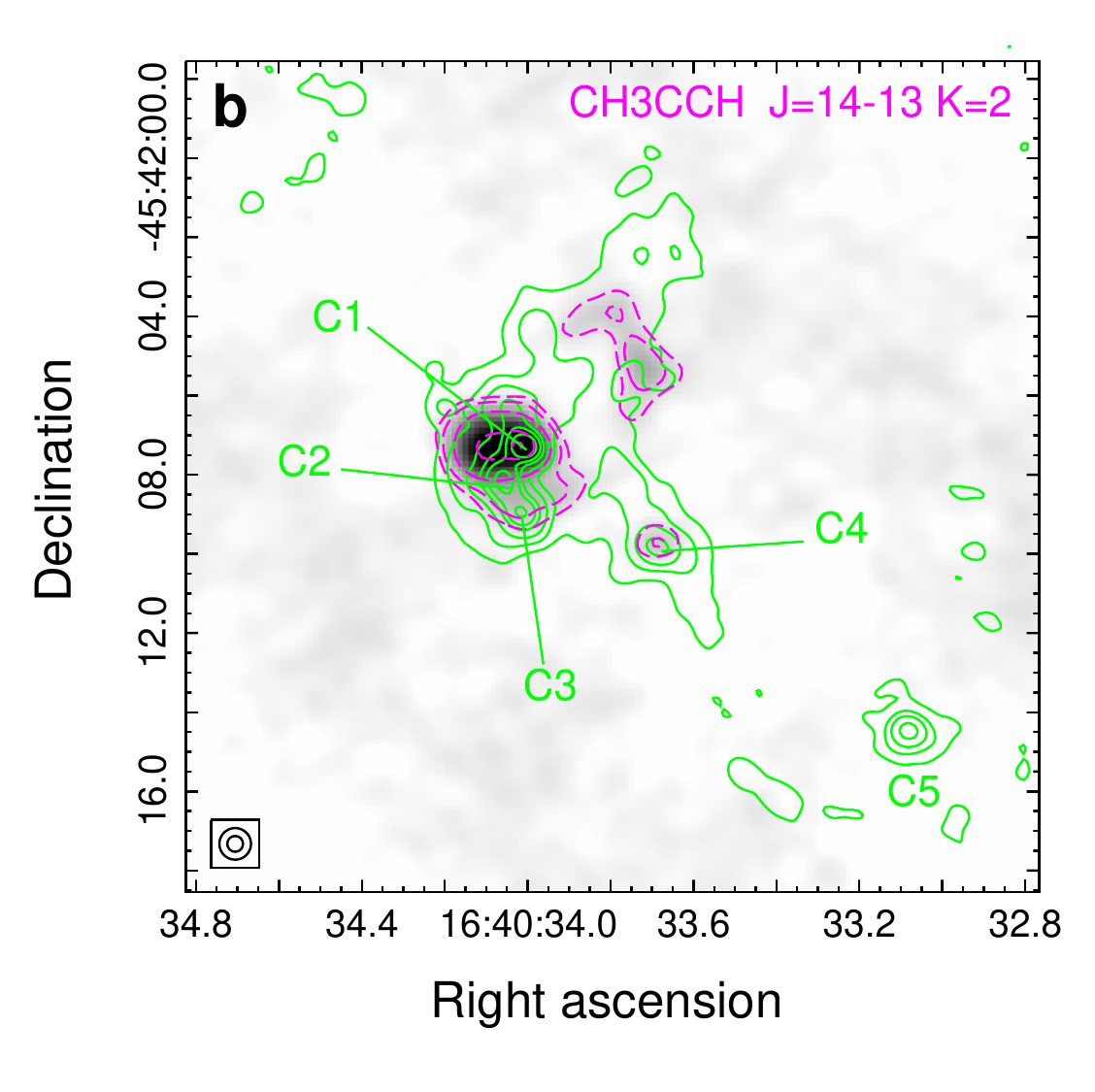}
\includegraphics[width=6.1cm]{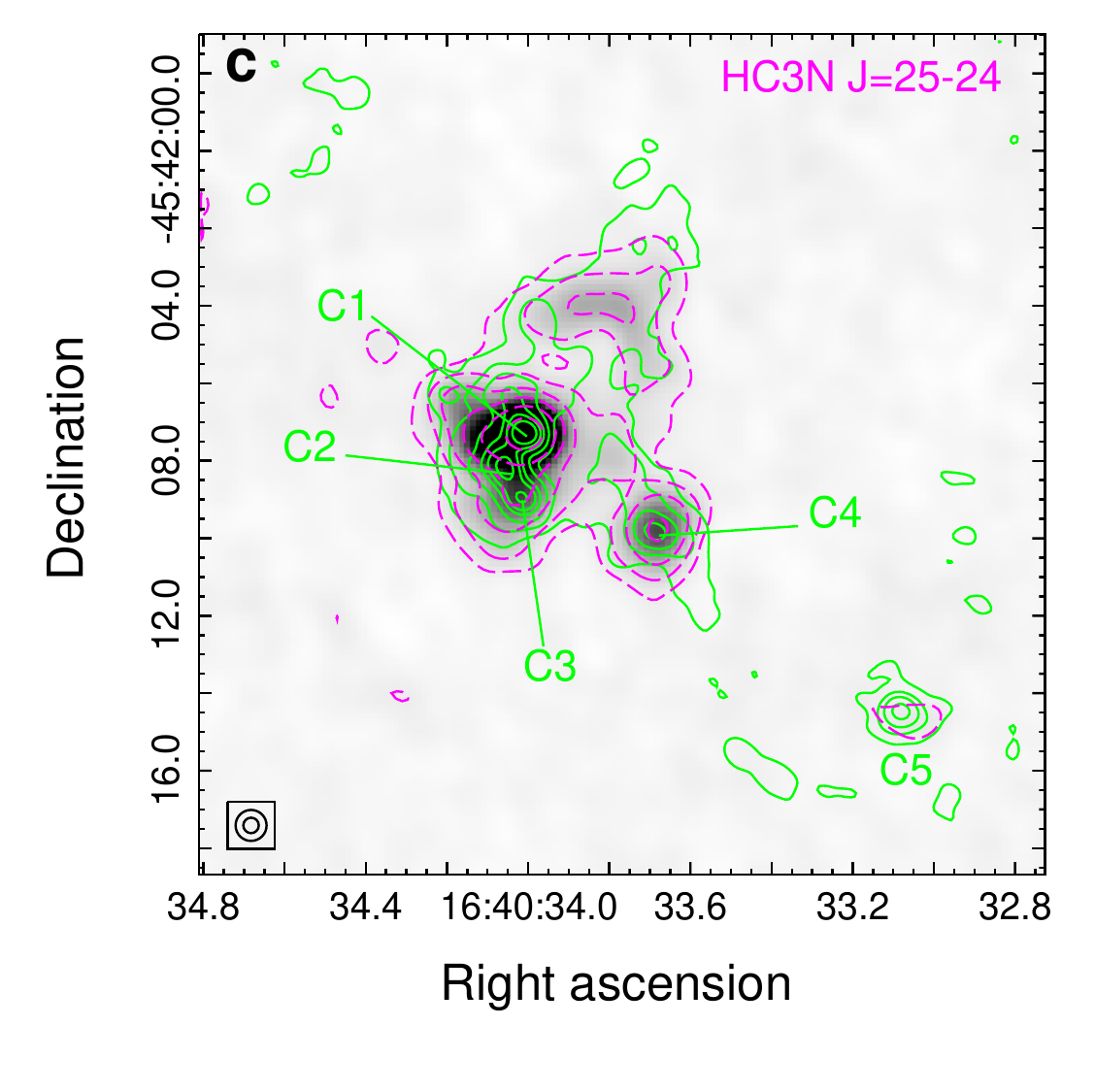} 
\includegraphics[width=6.3cm]{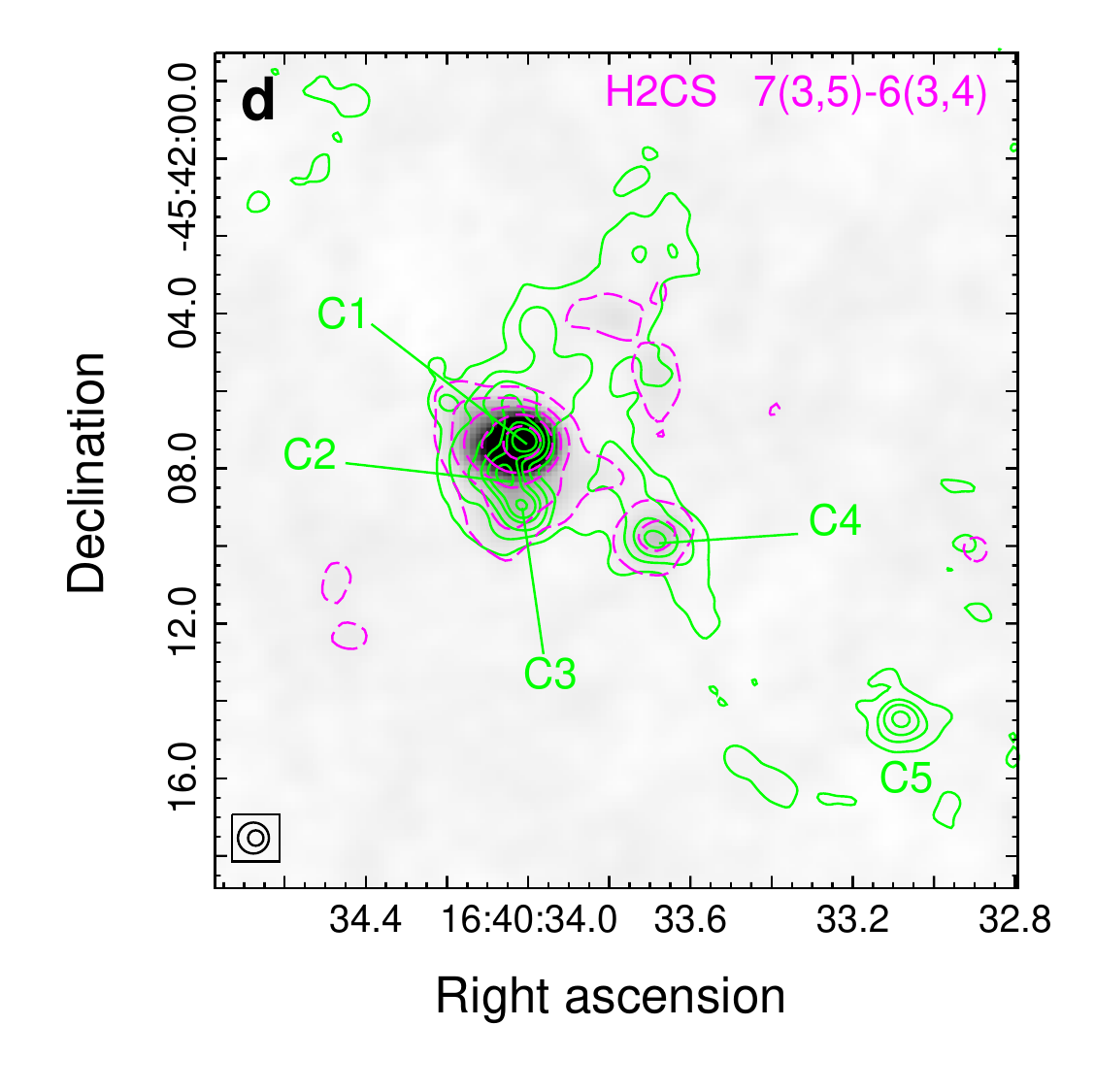}
\includegraphics[width=6.2cm]{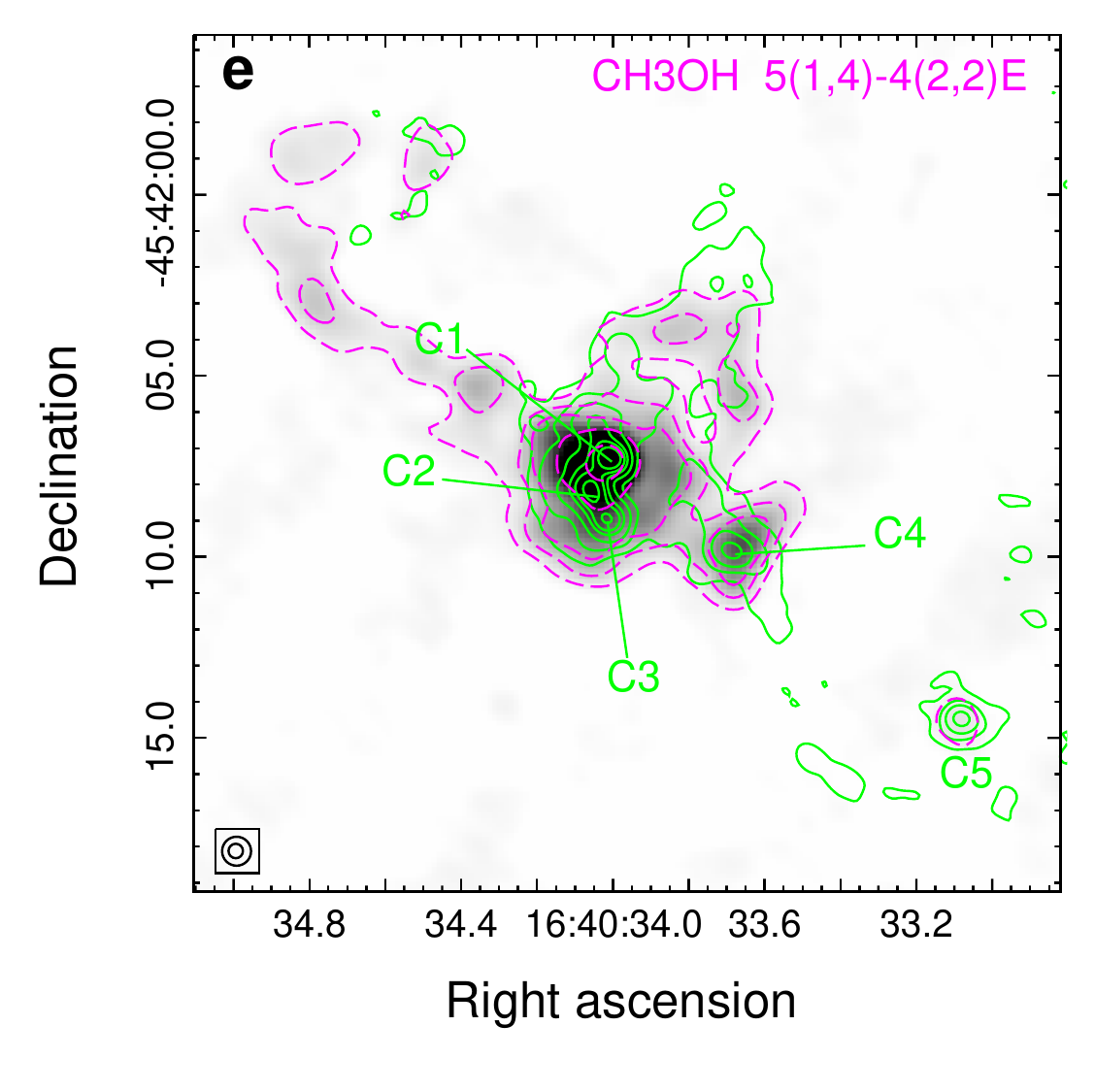}
\includegraphics[width=6.2cm]{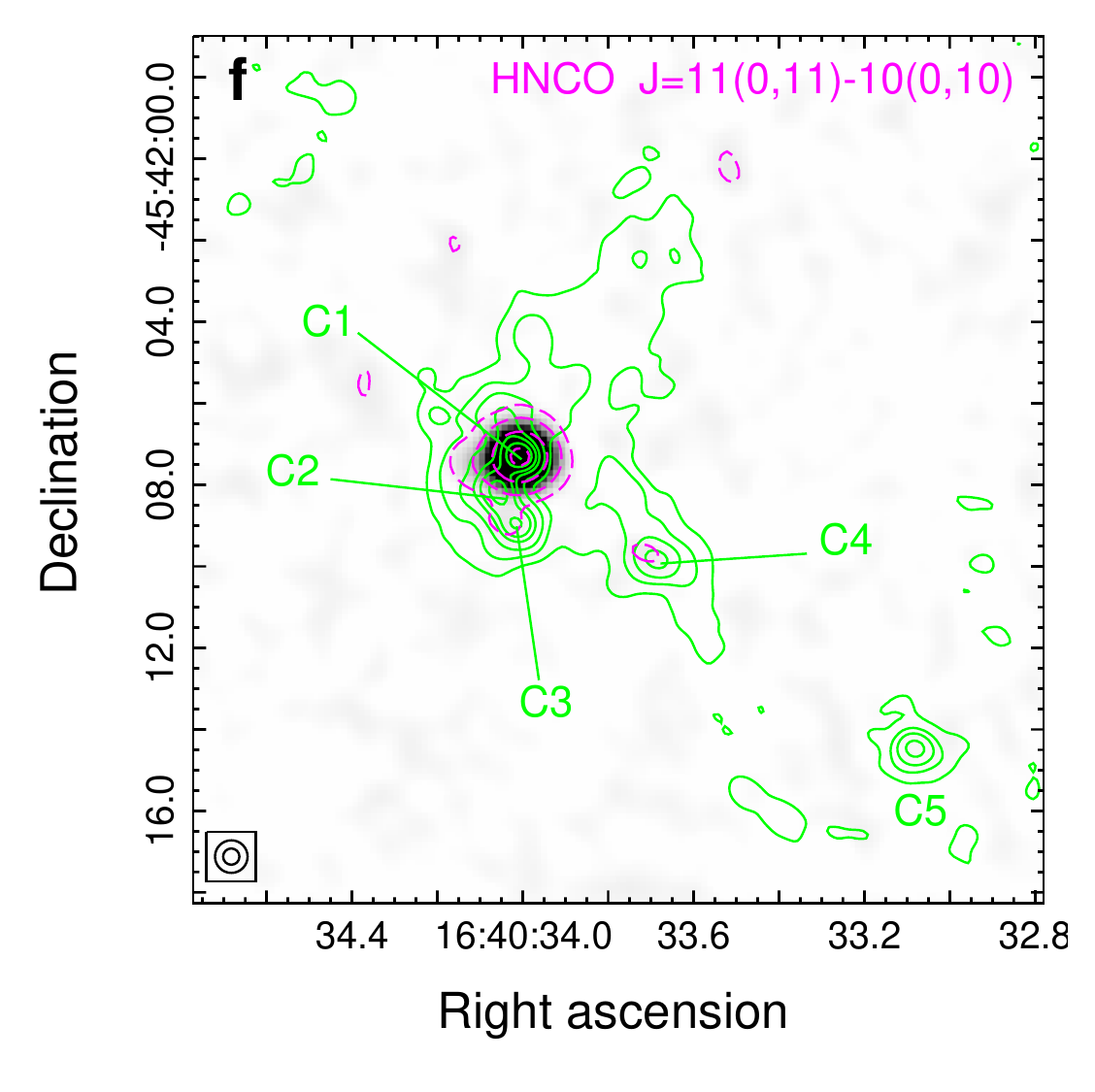}
\includegraphics[width=6.2cm]{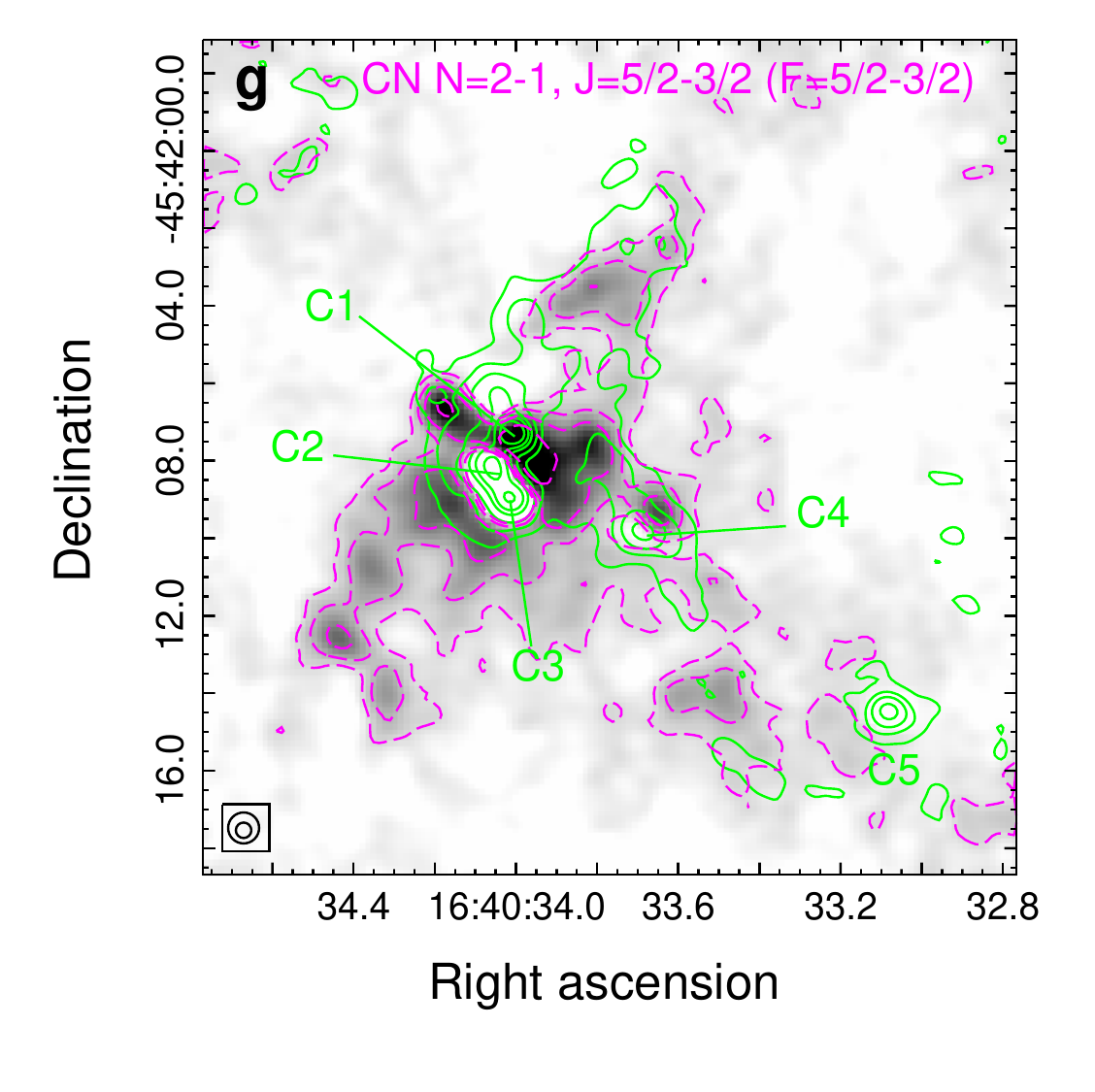}
\includegraphics[width=6.1cm]{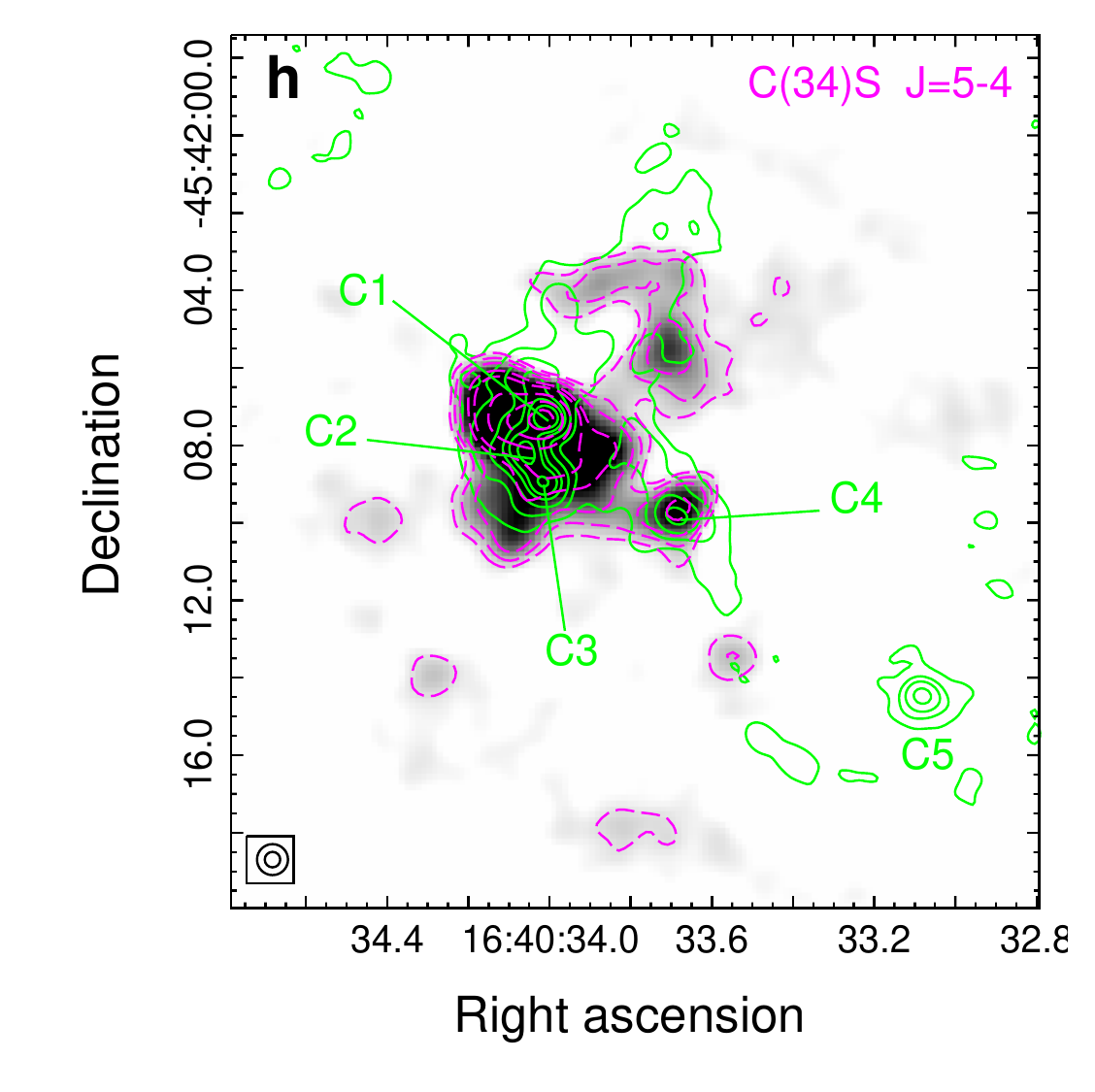}
\includegraphics[width=6.2cm]{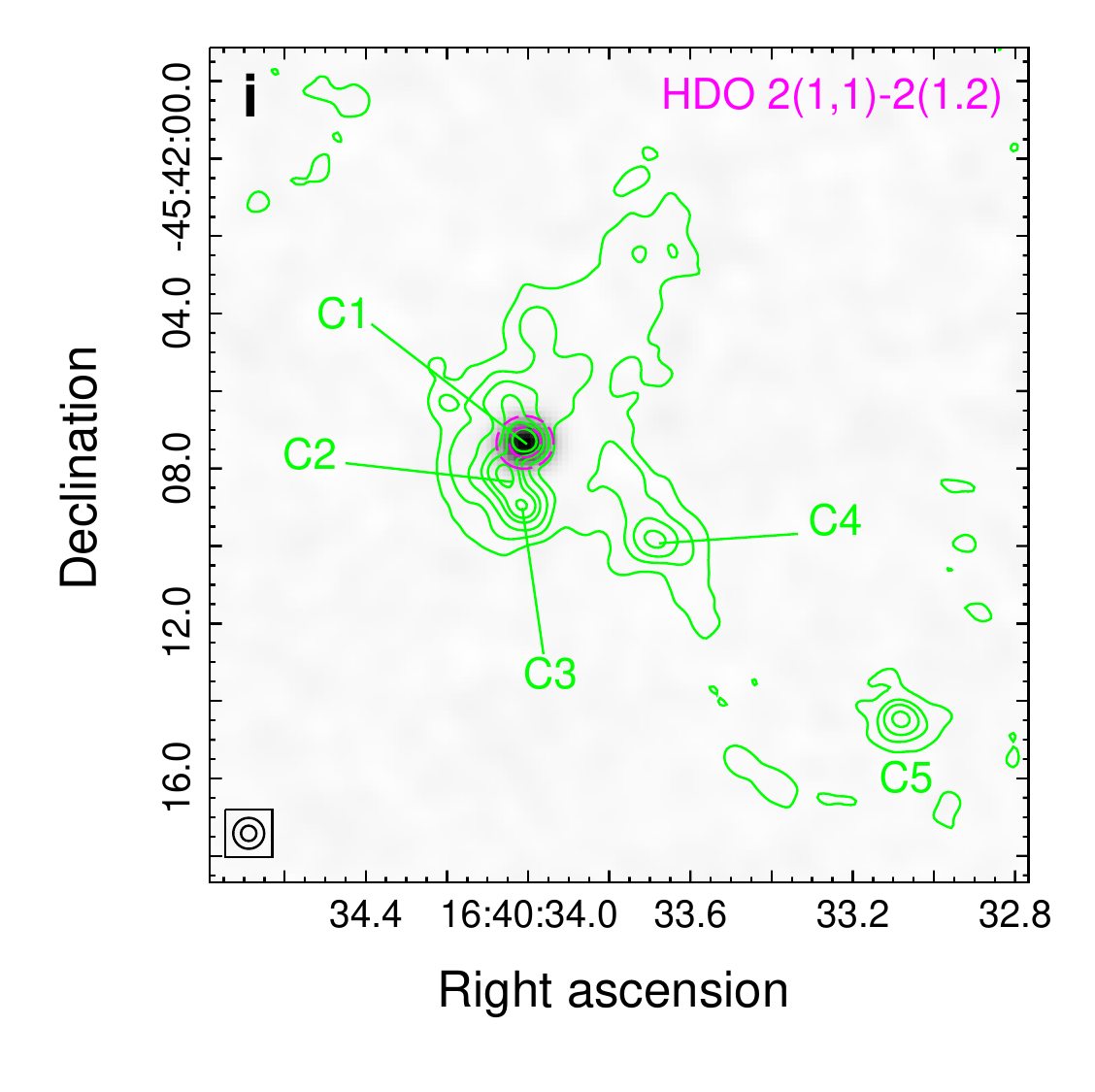}
\caption{Integrated emission maps (moment 0) of the selected molecular lines (indicated in each panel and in Table \ref{molecLines}). Dashed magenta contours represent the molecular line emission at levels as follows (in Jy beam$^{-1}$ \ks): 0.1, 0.3, 1, and 4 (CH$_{3}$CN), 0.05, 0.10, 0.20, 0.50, and 0.65 (CH$_3$CCH), 0.04, 0.1, 0.25, 0.6, 1.3, and 3 (HC$_3$N), 0.05, 0.2, 0.5, 1.2, and 3 (H$_{2}$CS), 0.1, 0.3, 0.6, 1.5, and 5 (CH$_{3}$OH), 0.05, 0.2, 1, and 3 (HNCO), 0.02, 0.04, 0.08, and 0.16 (CN),  0.07, 0.12, 0.2, 0.4, 0.6, 1.2, 2, and 3 (C$^{34}$S), and 0.5, 1, and 2 (HDO). The integration velocity interval ranges from $-$67 to $-$61~\ks~for all molecular lines, except for H$_{2}$CS and CH$_{3}$OH ($-$67 to $-$62~\ks) and for HNCO ($-$67 to $-$63~\ks). The green contours represent the continuum emission at 340~GHz with levels at 1, 10, 30, 60, 90, and 140~mJy beam$^{-1}$. The beams of 340~GHz continuum and line emissions are indicated at the bottom left corner.}
    \label{molecFig}
\end{figure*}

\subsubsection{CH$_3$CN and CH$_3$CCH: temperatures and column densities of the core C1}
\label{temperature}

Methyl cyanide (CH$_3$CN) and methyl acetylene (CH$_3$CCH) have been proven to be reliable tracers of physical conditions, such as temperature and density, and has been extensively studied towards several hot molecular cores \citep[e.g.,][and reference therein]{remijan2004,calcutt2019,brouillet2022,ortega2022}.
Their rotational transitions are characterized by two quantum numbers, namely, the total angular momentum (J) and its projection on the principal symmetry axis (K). Given that these molecules are top-symmetric rotors, present many K projections that are closely spaced in frequency, which favors its observation. Moreover, in both molecules, transitions with $\Delta K \neq 0$ are forbidden. Thus, the relative populations of different K-ladders are dictated only by collisions and as a result, CH$_3$CN and CH$_3$CCH act as excellent temperature probes.

In particular, given the small electric dipole moment of the CH$_3$CCH molecule ($\mu$ = 0.78~D), line thermalization occurs at densities as low as about 10$^4$~cm$^{-3}$ \citep[e.g.,][]{molinari2016}. 

Figure\,\ref{molecFig}-a and -b shows the moment 0 maps of the CH$_3$CN J=13--12 and CH$_3$CCH J=14--13 transitions, respectively, both at K=2 projection. The green contours represent the ALMA continuum emission at 340~GHz. The spatial distribution of both molecules overlaps with the position of cores C1 to C4, but while the CH$_3$CN peak positionally coincides with core C1, the emission peak of CH$_3$CCH appears shifted about a beam size relative to this core.  In particular, the methylacetylene emission  exhibits an arc-like structure towards the northeast in positional coincidence with the faint extended emission of the continuum at 340~GHz. 

Figure\,\ref{ch3cn.spectra} shows the CH$_3$CN J=13--12 (top panel) and CH$_3$CCH J=14--13 (bottom panel) spectra towards the core C1. It is important to mention that for the other cores, the CH$_3$CN and CH$_3$CCH spectra have no K projections above 5~$\sigma$ rms level from K$=$3, with which the temperature estimate is restricted to core C1. 

Table\,\ref{ch3cn_parameters} shows the tabulated parameters for all K projections of CH$_3$CN J=13--12 and CH$_3$CCH J=14--13 transitions detected towards core C1. Columns 1 and 2 show the K projection and the rest frequency, respectively, obtained from the NIST catalogue\footnote{https://physics.nist.gov/cgi-bin/micro/table5/start.pl}. Column 3 presents the upper energy level (${\rm E_{u}/k}$) extracted from the LAMDA database\footnote{https://home.strw.leidenuniv.nl/$~$moldata/}, and Column 4 shows the line strength of the projection multiplied by the dipole moment of the molecule (${\rm S_{ul}\mu^{2}}$).

 \begin{figure}[h]
   \centering
   \includegraphics[width=9cm]{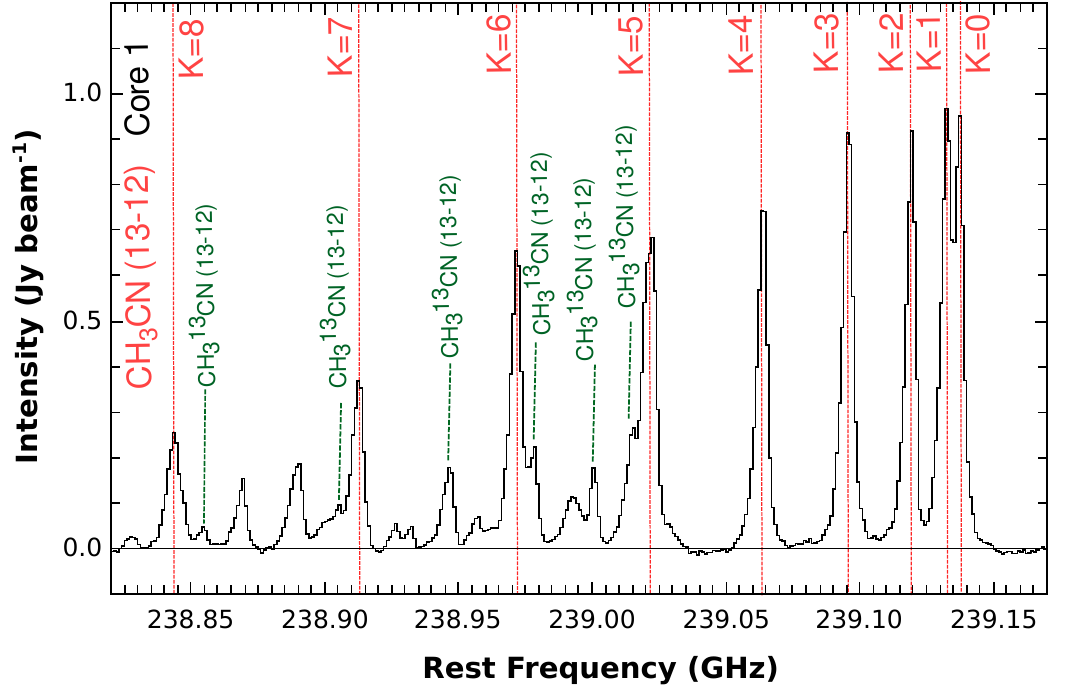}
   \includegraphics[width=9.7cm]{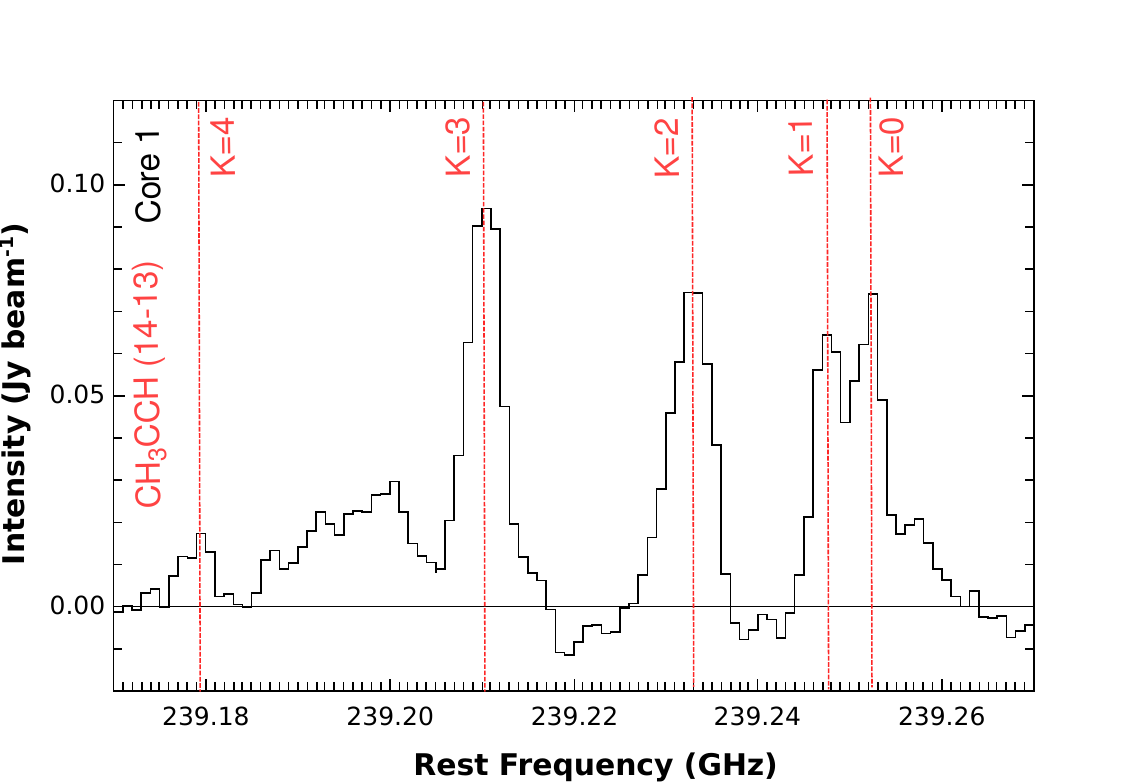}
    \caption{Average spectra of the temperature tracer molecules taken over the full size of core C1. Top panel: CH$_3$CN J=13--12. Bottom panel: CH$_3$CCH J=14--13. The K projections are indicated.}
    \label{ch3cn.spectra}
\end{figure}

Table\,\ref{ch3cn_fitting} shows the main parameters derived from the  Gaussian fittings to the CH$_3$CN and CH$_3$CCH spectra towards the core C1. Columns 2, 3, 4, and 5 show the peak intensity, the central velocity (${\rm v_c}$), the FWHM  ($\Delta$v), and the integrated intensity (W), respectively. The integrated intensities were used to construct the rotational diagram (RD) presented in Fig.\,\ref{RotDiagr_CH3CN_core1}. Thus, using the RD analysis \citep[][and references therein]{goldsmith99} and assuming LTE conditions, optically thin lines, and a beam filling factor equal to the unity, we can estimate the rotational temperatures (${\rm T_{rot}}$) and the column densities for the core C1 using both molecules. This analysis is based on a derivation of the Boltzmann equation,

\begin{equation}
{\rm ln\left(\frac{N_u}{g_u}\right)={\rm ln}\left(\frac{N_{tot}}{Q_{rot}}\right)-\frac{E_u}{kT_{rot}}},   \label{RotDia} 
\end{equation}

\noindent where ${\rm N_u}$ represents the molecular column density of the upper level of the transition, ${\rm g_u}$  the total degeneracy of the upper level, ${\rm N_{tot}}$ the total column density of the molecule, ${\rm Q_{rot}}$ the rotational partition function, and k the Boltzmann constant. 

Following \citet{miao95}, for interferometric observations, the left-hand side of Eq.\,\ref{RotDia} can be estimated from:

\begin{equation}
{\rm ln\left(\frac{N_u^{obs}}{g_u}\right)={\rm ln}\left(\frac{2.04 \times 10^{20}}{\theta_a \theta_b}\frac{W}{g_kg_l{\nu_0}^3S_{ul}{\mu_0}^2}\right)},   
\label{RD} 
\end{equation}

\noindent where ${\rm N_u^{obs}}$ (in cm$^{-2}$) is the observed column density of the molecule under the above mentioned conditions, $\theta_a$ and $\theta_b$ (in arcsec) are the major and minor axes of the clean beam, respectively, W (in Jy beam$^{-1}$ \ks) is the integrated intensity of each K projection, ${\rm g_k}$ is the K-ladder degeneracy, ${\rm g_l}$ is the degeneracy due to the nuclear spin, ${\rm \nu_0}$ (in GHz) is the rest frequency of the transition, ${\rm S_{ul}}$ is the line strength of the transition, and $\mu_0$ (in Debye) is the permanent dipole moment of the molecule.  The free parameters, (${\rm N_{tot}/Q_{rot}}$) and ${\rm T_{rot}}$ were determined by a linear fitting to Eq.\,\ref{RotDia} (see Fig.\ref{RotDiagr_CH3CN_core1}). Finally, using the tabulated value for ${\rm Q_{rot}}$ at the corresponding temperature, extracted from the CDMS database\footnote{https://cdms.astro.uni-koeln.de/cdms/portal/queryForm}, we obtain the CH$_3$CN and CH$_3$CCH column densities for core C1 (see Table\,\ref{ch3cn_DR_parameters}). The CH$_{3}$CN K=5 to K=7 projections are blended with some CH$_{3}^{13}$CN isotopologue projections (see Fig. \ref{ch3cn.spectra}-top panel). Therefore, in such cases two Gaussian components were fitted. 

The CH$_3$CN K=7 and K=8 components show a central velocity shift of about 1~\ks~with respect to the systemic velocity, which suggests possible contamination of other lines. In fact, the rotational diagram with the measured W for these K projections yields a temperature above 500 K, which seems to be too high for the gas traced by this molecular species. Thus, we looked for potential contamination lines into the Splatalogue platform (JPL and CDMS databases). We did not find any obvious contamination line beyond several rare complex molecules with very weak intensities. However, it is likely that in such line-rich spectra (see Sect.\,\ref{appendix}) there could be still unidentified lines, probably from vibrational or torsional states of known molecules. Therefore, assuming that 50 percent of the components area for K=7 and K=8 comes from contamination of unidentified lines, we use half of the integrated intensities values for these projections to build the rotational diagram.

Optical depths effects, that could result in an underestimation of the intensity of a line, tend to be more noticeable in lower projections. This would produce a flattening of the slope in a RD graphic, leading to anomalously large values for $T_{rot}$. The method, proposed by \citet{goldsmith99}, iteratively correct individual $\rm {N_u/g_u}$ values by multiplying by the optical depth correction factor, $C_{\tau}=\tau/(1-e^{-\tau})$. However, we find that the $\tau$ corresponding to K=0 projection is lower than 0.06 for both molecules, which leads to a correction factor less than 3 per cent, and therefore, the rotation temperature would not be overestimated.

\begin{table}
\caption{Tabulated parameters for all the K projections of CH$_{3}$CN (13--12) and CH$_{3}$CCH (14--13) lines detected above 5~$\sigma$ noise level towards the core C1.}
\centering
\begin{tabular}{c c c c }
\hline\hline
    Proj.     &   Rest frequency &  ${\rm E_u/k}$ & ${\rm S_{ul}\mu^2}$\\
   K       &   (GHz)   & (K)       &   (Debye$^2$) \\ 
 \hline                 
  &     &\hspace{-1.5cm} CH$_{3}$CN J=13--12   &  \\
 \hline
    0     &  239.137    &  80.3 & 199.1\\
    1     &  239.133    &  87.5 & 198.6\\
    2     &  239.119    & 108.9 & 195.0\\
    3     &  239.096    & 144.6 & 188.5\\
    4     &  239.064    & 194.6 & 180.8\\ 
    5     &  239.022    & 258.9 & 170.2\\
    6     &  238.972    & 337.4 & 157.2\\
    7     &  238.912    & 430.1 & 141.8\\
    8     &  238.843    & 537.0 & 124.1\\
 \hline                 
  &     &\hspace{-1.5cm} CH$_{3}$CCH J=14--13   &  \\
 \hline
    0     &  239.252    &  86.1 & 7.9\\
    1     &  239.247    &  93.3 & 7.8\\
    2     &  239.234    & 115.5 & 7.7\\
    3     &  239.211    & 151.1 & 7.5\\
    4     &  239.179    & 201.7 & 7.2\\  
 \hline
\end{tabular}
\label{ch3cn_parameters}
\end{table}

\begin{table}
\small
\caption{Gaussian fittings parameters for the CH$_{3}$CN (13--12) and CH$_{3}$CCH (14--13) K projections detected above 5~$\sigma$ noise level from the spectral of Fig. \ref{ch3cn.spectra}.}
\centering
\begin{tabular}{c c c c c}
\hline
\hline
    Proj.     &    Peak Int.   & v$_{c}$ & $\Delta$v &          W           \\
      K       & (Jy beam$^{-1}$)  &  \ks &  \ks    & (Jy beam$^{-1}~\times$ \\    
              &                &         &         & $\times$ \ks \\
\hline                 
& {\bf CH$_{3}$CN} & {\bf J=13--12} & & \\
\hline
    0     &  1.03$\pm$0.12 & $-64.2\pm0.9$ & 4.8$\pm$1.2 & 5.91$\pm$1.47 \\
    1     &  1.00$\pm$0.14 & $-64.3\pm0.9$ & 4.7$\pm$1.3 & 6.33$\pm$1.51 \\
    2     &  0.95$\pm$0.12 & $-64.2\pm0.6$ & 4.5$\pm$0.6 & 5.12$\pm$1.33 \\
    3     &  0.98$\pm$0.13 & $-63.9\pm0.6$ & 4.4$\pm$0.7 & 5.52$\pm$1.43 \\
    4     &  0.78$\pm$0.08 & $-64.2\pm0.6$ & 4.4$\pm$0.6 & 4.54$\pm$1.04 \\
    5     &  0.71$\pm$0.06 & $-64.2\pm0.7$ & 4.7$\pm$0.7 & 4.35$\pm$0.96 \\
    6     &  0.68$\pm$0.09 & $-64.2\pm0.7$ & 4.2$\pm$0.7 & 3.81$\pm$1.02 \\ 
    7     &  0.38$\pm$0.04 & $-65.4\pm0.7$$^{*}$ & 3.8$\pm$0.8 & 1.93$\pm$0.54 \\
    8     &  0.28$\pm$0.04 & $-65.5\pm0.7$$^{*}$ & 4.5$\pm$0.5 & 1.76$\pm$0.42 \\ 
\hline                 
& {\bf CH$_{3}$CCH} & {\bf J=14--13} & & \\
\hline
    0     &  0.085$\pm$0.013 & $-63.9\pm1.3$ & 3.3$\pm$1.4 & 0.57$\pm$0.21 \\
    1     &  0.079$\pm$0.012 & $-64.2\pm1.2$ & 3.1$\pm$1.2 & 0.37$\pm$0.14 \\
    2     &  0.091$\pm$0.012 & $-63.7\pm0.6$ & 3.6$\pm$0.4 & 0.46$\pm$0.17 \\
    3     &  0.115$\pm$0.014 & $-63.9\pm0.7$ & 3.5$\pm$0.5 & 0.54$\pm$0.19 \\
    4     &  0.018$\pm$0.004 & $-63.9\pm0.7$ & 2.9$\pm$0.4 & 0.07$\pm$0.02 \\
\hline
\multicolumn{5}{l}{*The v$_c$ of these projections suggest contamination of other lines. }
\end{tabular}
\label{ch3cn_fitting}
\end{table}

\begin{figure}[h]
   \centering
   \includegraphics[width=9cm]{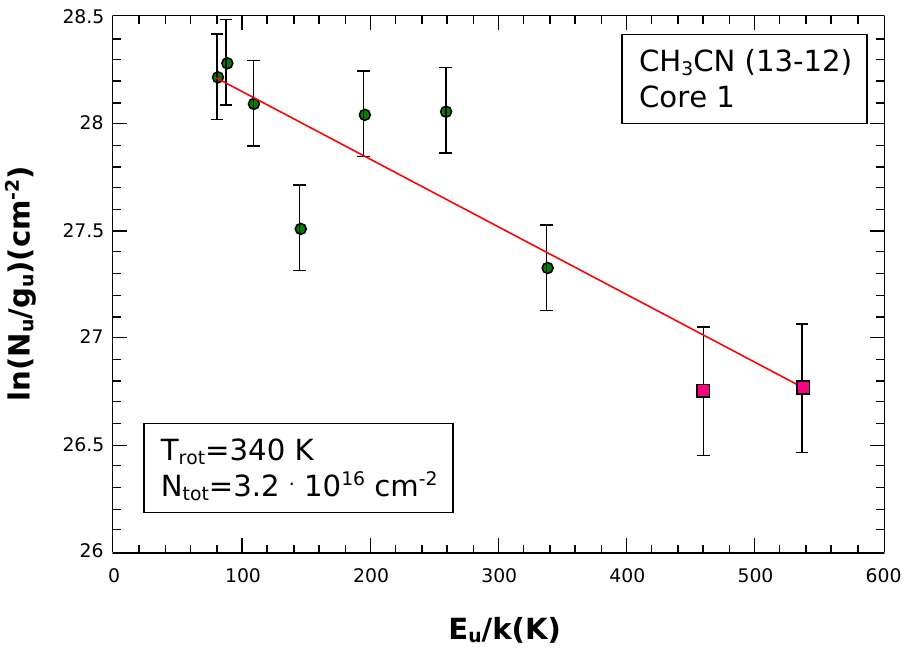}
   \includegraphics[width=9cm]{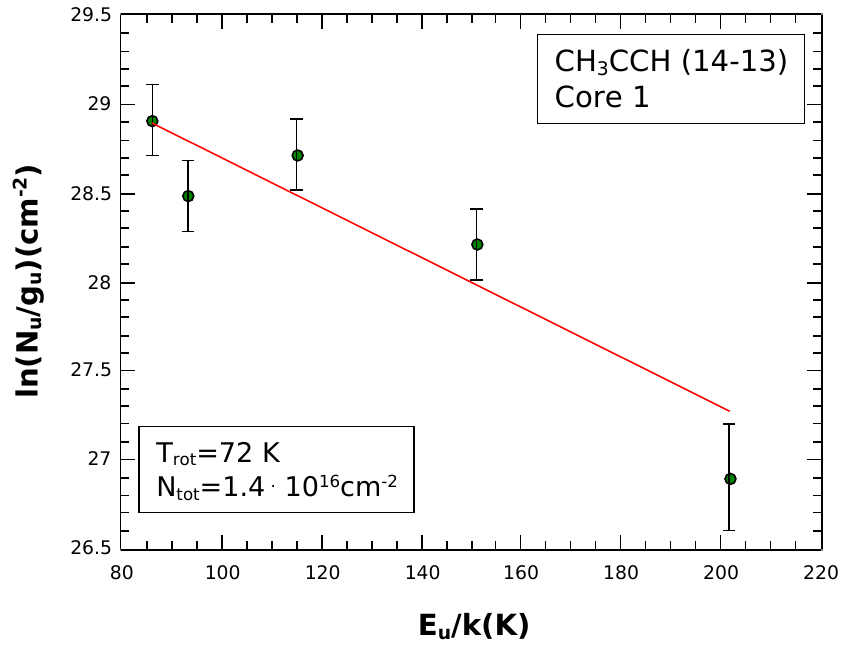}
   \caption{Rotational diagrams for the core C1. Top panel: CH$_3$CN J=13--12. Bottom panel: CH$_3$CCH J=14--13. The red lines show the best linear fitting of the data. The magenta squares in top panel correspond to CH$_3$CN J=13--12 K=7 and K=8 projections, for which half of the integrated intensities values were considered.}
   
    \label{RotDiagr_CH3CN_core1}
\end{figure}

\begin{table}
\caption{Rotational temperature, and molecular column density for the core C1 derived from a rotational diagram analysis using the CH$_{3}$CN J=13--12 and CH$_{3}$CCH J=14--13 transitions.}
\centering
\begin{tabular}{c c c }
\hline\hline
Molecule & T$_{\rm rot}$ &  Column density     \\
     & (K)       & $\times$10$^{16}$(cm$^{-2}$)   \\ 
   
\hline                 
 CH$_{3}$CN       & 340$\pm$95 & 3.2$\pm$1.1  \\
 CH$_{3}$CCH      &  72$\pm$13 & 1.4$\pm$0.5  \\
 
\hline
\hline
\end{tabular}
\label{ch3cn_DR_parameters}
\end{table}

\subsubsection{Mass and kinematic of the core C1}
\label{masses}

We estimate the mass of the core C1 from the continuum emission at 340~GHz using each rotational temperature derived in the previous section. 

Considering that, at the early evolutionary stage of AGAL\,338, the contribution of free-free continuum emission at 340~GHz is negligible (e.g. \citealt{isequilla21}), it is reasonable to assume that, at this frequency,  the submillimeter continuum is mainly tracing the dust emission. 
Then, the mass of gas of the core C1 (see Table\,\ref{ch3cn_DR_parameters}) was estimated from the dust continuum emission at 340~GHz ($\lambda \sim$ 0.9~mm) following  \citet{kau08},

\begin{eqnarray}
{\rm M_{gas}=0.12~M_\odot \left[exp\left(\frac{1.439} {(\lambda/mm)(T_{dust}/10~K)}\right)-1\right]} \\ \nonumber {\rm \times\left(\frac{\kappa_{\nu}}{0.01~cm^2~g^{-1}}\right)^{-1}\left(\frac{S_{\nu}}{Jy}\right)\left(\frac{d}{100~pc}\right)^2\left(\frac{\lambda}{mm}\right)^3}
\label{massdust}
\end{eqnarray}

\noindent where ${\rm T_{dust}}$ is the dust temperature and $\kappa_{\nu}$ is the dust opacity per gram of matter at 870~$\mu$m, for which we adopt the value of 0.0185~cm$^2$g$^{-1}$ \citep[][and references therein]{csengeri2017a}. We assume thermal coupling between dust and gas (T$_{\rm dust}$=T$_{\rm kin}$), where T$_{\rm kin}$=T$_{\rm rot}$. 
 
Anyway we use a T$_{\rm kin}$ ranging from a typical desorption temperature in hot cores of about 120 K (e.g. \citealt{busch22}) to the temperature estimated from the rotational diagram of the CH$_3$CN (see Sect.\,\ref{molecular}), to obtain a core mass ranging from 3 to 10 \msol.

Despite the fact that estimating mass for cores based on dust emission is the most reliable method, there are some sources of uncertainty. Considering an absolute flux uncertainty $\leq$ 10\% for ALMA observations in band 7, a dust temperature uncertainty of about 20\% and a distance uncertainty of $\sim$10\%, the mass uncertainty would be about 50\%.

Figure\,\ref{CH3CN_mom1_k4} shows the CH$_{3}$CN J=13--12 moment 1 map for the K=4 projection integrated between $-$66 and $-$62~\ks. It can be appreciated that, at this K-projection,  the emission is concentrated only towards the core C1. The gas related to this core exhibits a clear velocity gradient perpendicular to the molecular outflow direction. This velocity gradient has been interpreted in several works as evidence of a rotating disk (e.g. \citealt{louvet16,ortega2022}). It is important to mention that this signature of disk rotation has only been found in this molecular species.

\begin{figure}[h]
   \centering
   \includegraphics[width=9cm]{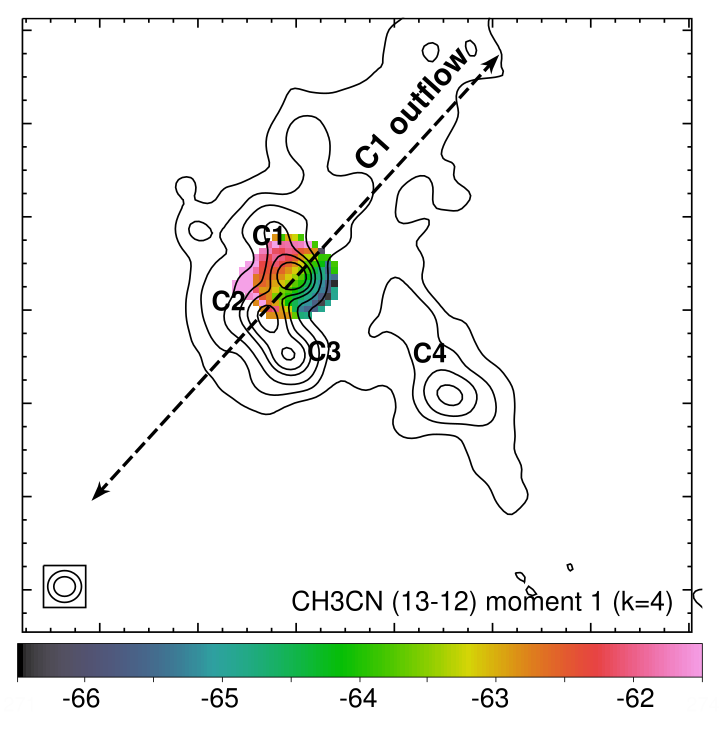}
    \caption{CH$_3$CN J=13--12 moment 1 map (from K=4 projection). The black contours represent the continuum emission at 340~GHz with levels at 1, 10, 30, 60, 90, and 140~mJy beam$^{-1}$. The dashed black line indicates the direction of the molecular outflow related to core C1. The beams are indicated in the bottom left corner.}
    \label{CH3CN_mom1_k4}
\end{figure}

\section{Discussion}

In this section, we discuss the implications of our findings and their potential impact on our understanding of high-mass star formation. We also compare our results with previous studies and highlight the most significant differences and similarities. Finally, we discuss the limitations of our study and suggest avenues for future research.

\subsection{A massive clump fragmented into low-intermediate mass cores?} 
\label{tempmass}

Massive clumps usually have a relatively low thermal Jeans mass, which predicts a high level of fragmentation.  \citet{csengeri2014} estimated an integrated flux of about 7.36~Jy at 870~$\mu$m for AGAL\,338. Assuming a typical temperature of 20~K, and considering a radius of the clump of 0.3~pc, we derive a  clump mass of $\sim$1260~\msol~ and a Jeans mass of $\sim$4~\msol, which suggest that AGAL\,338 would be unstable to fragmentation. Specifically, based on the dust continuum emission at 340~GHz, we found that the fragmentation of AGAL\,338 gave rise to at least five molecular cores, labeled from C1 to C5. 

Although the CH$_3$CCH and CH$_3$CN molecules were detected towards most of the cores, it was only possible to estimate temperatures for core C1. Using the CH$_3$CCH and CH$_3$CN molecules, we derived temperatures of about 72 and 340~K, respectively. In particular, a temperature value of 340~K is among the highest temperature values found by \citet{her14} towards the compact component of several hot cores. The detection above 5$\sigma$ noise level of the CH$_3$CN J=13--12 K=8 projection (E$_u$=537~K),  the presence of several CH$_{3}$CN v$_{8}$=1 lines (see Fig. \ref{band6spw0}), and the richness of the spectra in the four spectral windows, suggest that a high temperature gas component is present in core C1. However, precisely due to the line richness of the spectra, and the contamination in the lines that this entails, it is very difficult to accurately estimate this high temperature value.

The discrepancy in the temperatures derived from CH$_{3}$CCH and CH$_{3}$CN can be indicating that each molecule is tracing different gas layers associated with the hot core. According to this, \citet{andron18} constrained the origin of these two molecular species in the envelope of a low-mass protostar and studied a chemical model that predicted the desorption of the CH$_3$CN molecule from the dust grains in a radii closer to the protostar (at higher temperatures) than for the CH$_3$CCH. This would explain the higher temperature value found for the core C1 from the methyl cyanide molecule. 

Using a range of temperatures going from 120 K (about the typical molecular desorption temperature in hot cores) to 340~K obtained from the rotational diagram of the CH$_3$CN, the mass of core C1 ranges from 3 to 10 \msol. The mass of such a core, the brightest and more active core embedded in AGAL\,338, is quite below the limit for a massive core (a few tens of solar masses) candidate to form high-mass stars in a scenario of monolithic collapse. In such a scenario, following \citet{duarte13}, who indicate that the efficiency for the core mass being converted to stellar mass is about 50\%, core C1 would give rise to a low-mass star.

It is not unreasonable to assume that fragmentation of the molecular clump AGAL\,338 seems to have produced  low- and/or intermediate-mass cores. Therefore, the only path for the formation of massive stars in this region should be cores acquiring mass through gas infalling from their parent structures. In other words, a competitive accretion scenario. However, although we have searched for signatures of converging gas filaments through a kinematic analysis of the gas in all molecules, we did not find any evidence of streams of gas feeding the cores.

\subsection{EGO\,G338 and the core molecular outflow activity}

We discuss the molecular outflow activity associated with the cores C1 and C2 in relation with the presence of the EGO\,G338, one of the brightest in the \citet{cyga2008}'s catalog, taking into account that an EGO is a MYSO candidate to produce molecular outflows.

In Sect.\,\ref{12CO}, we characterize the outflow activity related to the cores C1 and C2. As shown in Fig.\,\ref{co-outflows}, while the red lobes associated with the cores C1 and C2 are spatially separated and relatively well collimated, the inner region between the blue lobes shows extended emission that connects them with a cone-like shape structure that opens towards the south. It is likely that this morphology is due to the presence of core C3, and the core C2 itself, which could be scattering the gas of both lobes. This could be a case of an interaction between molecular outflows and dense cores as it was found in the OMC-2 region \citep{shima08,sato22}.

The near-IR counterpart of the molecular outflow activity manifests as extended emission arising from the core C1 and pointing towards the southeast direction, which perfectly match the position of the outflow blue-OC1 (see Fig.\,\ref{VISTA_Ks}). It is well known that the origin of the continuum emission at K${\rm _s}$-band around protostars can be explained as a scattered light nebulosity, where the light scattering process occurs in the walls of a cavity that was cleared out in the circumstellar material by a jet \citep{bik2006} and/or emission of H$_2$ likely associated with shocked gas \citep[][and refences therein]{mccoey2004}. 
Interestingly, the cavity/jet nebula, as observed towards other similar sources \citep[see][and references therein]{wei06,paron2016}, extends only to one side. To justify this unidirectional asymmetry, it was proposed that the observed near-IR features might be related to a blue-shifted jet with the red-shifted counterpart not detected at the near-IR bands because they are more highly extinct. This phenomenon is clearly manifested in this source.

Regarding the mid-IR emission, a comparison between Figs.\,\ref{intro} and\,\ref{co-outflows} clearly shows that the 4.5~$\mu$m extended emission of the EGO\,G338 and the outflow blue-OC1 positionally coincide and exhibit the same inclination in the plane of the sky. This suggests that the main contribution to the EGO emission comes from the outflow activity of the core C1, in particular, from the blue lobe. As it occurs with the near-IR emission, the mid-IR counterpart associated with the outflow red-OC1 is not detected.

In what follows, we perform a comparison between some of our main results regarding molecular outflow activity and the work of \citet{li2020}, which is one of the most comprehensive and modern studies carried out with similar observations as used in this work. The authors present a statistical study with ALMA data (beam $\sim$ 1$\farcs$2) towards more than 40 dense cores with associated molecular outflow activity. 

We estimated molecular outflows masses ranging from 0.08 to 0.77~\msol~(see Table\,\ref{outflow_parameters}) and dynamical ages on the order of 10$^{3}$~yrs. \citet{li2020} found outflow masses ranging from 0.001 to 0.32~\msol~and dynamical ages going from about 10$^3$ to some 10$^4$~yr, which indicates that the outflows found towards AGAL\,338 are among the youngest and most massive ones.

\citet{li2020} found a median ratio of outflow mass to core mass of about $8 \times 10^{-3}$. Thus, considering that the range of masses estimated for the outflow red-OC1 goes from 0.25 to 0.77~\msol, we conclude that a  mass of about 10~\msol~for the core C1 would be more likely than the lower value of about 2~\msol.

We estimated energies for the molecular outflow red-OC1 that goes from $3.9 \times 10^{45}$ to $1.2 \times 10^{46}$~erg, while \citet{li2020} found energies for the molecular outflows of the sample ranging from $4.0 \times 10^{41}$ to $1.2 \times 10^{45}$~erg. Therefore the molecular outflow red-OC1 is at least three times more energetic than the most energetic outflow of the \citet{li2020}'s work. 

Finally, following \citet{li2020}, we estimate for the core C1 a mass accretion rate, ${\rm \dot{M}}$, that goes from $1.5 \times 10^{-5}$ to $4.2 \times 10^{-5}$~\msol~yr$^{-1}$. Thus, using the highest value for ${\rm \dot{M}}$, and considering the estimated dynamical age of about $4.2 \times 10^3$~yr for the outflow red-OC1, the young protostar embedded in the core C1 could have at most 0.4~\msol. Therefore, even assuming the highest value of about 10~\msol~for the mass of the core C1, we consider that a high-mass star is unlikely to form in this core.

\subsection{Chemistry} 

Star-forming regions, and in particular HMCs, are excellent astrochemical laboratories to understand how complex molecules are formed in space (e.g.; \citealt{jorgen20,coletta20}). In turn, this understanding helps us to better characterize these interesting condensations of gas and dust where the stars form.
 
 In this section we discuss the presence of the molecules presented in Fig.\,\ref{molecFig} in the light of the most current astrochemical knowledge. Such molecular species are discussed individually: the morphology of the emission in the whole investigated region, chemical and physical conditions that they trace, etc., in order to obtain a comprehensive chemical interpretation of the analyzed molecular cores in terms of the star formation processes. Regarding the core C1, 
 in Appendix\,A we present the spectra of the four ALMA band\,6 spectral windows, which show the chemical richness and complexity of this core, the main one.

\subsubsection{CH$_{3}$CN and CH$_{3}$CCH}  

Propyne (also called methyl acetylene, CH$_3$CCH) and methyl cyanide (CH$_3$CN) are symmetric top molecules used as good indicators of temperature (e.g. \citealt{brouillet2022}, and see Sect.\,\ref{temperature}). These molecular species are usually detected in hot molecular cores (e.g. \citealt{brouillet2022}) and CH$_3$CN was also found in protoplanetary disks \citep{oberg15}. 
The CH$_{3}$CCH is likely produced in interstellar ices through combination of radicals \citep{kalenskii} and via successive hydrogenation of physisorbed C$_{3}$ \citep{hickson16,wong18}. CH$_{3}$CN is also formed in interstellar grains through radicals recombination such as CN$\cdot$ and CH$_{3}\cdot$ \citep{her14}. However, as different authors point out \citep{andron18,brouillet2022}, CH$_{3}$CN would trace the inner regions of the cores because it needs a higher temperature to sublimate from dust grains surface, while CH$_{3}$CCH emission is found preferentially tracing the colder envelopes, which would explain the discrepancy in the temperature obtained from both emissions. By inspecting Fig.\,\ref{molecFig} (a) and (b), the morphology of the emission of both molecules is quite similar. They are concentrated mainly at the bulk of the emission that contains cores C1, C2, and C3. Core C4 also presents emission of both molecules. In the case of the CH$_{3}$CCH, a feature also appears towards the northwest in correspondence with the position of the outflow red-OC1 (see Fig.\,\ref{co-outflows}). This may suggest that the outflow activity in this region could desorb molecular species frozen in the dust grains enriching the gas phase chemistry in the diffuse gas, for instance releasing  CH$_{4}$, which seems to be important for the CH$_{3}$CCH chemistry in the gas phase \citep{calcutt2019}.

\subsubsection{HC$_{3}$N}
\label{hc3n}

It is known that the shortest cyanopolyyne HC$_{3}$N, the cyanoacetylene, is helpful to explore gas associated with hot molecular cores 
\citep{bergin96,taniguchi16,duronea19}. As shown in Fig.\,\ref{molecFig} (c), the HC$_{3}$N emission is mainly concentrated in the cores, and in general encompasses the continuum emission. Towards the north-west it is observed another maximum of the emission of this molecular species not associated with any core traced in the continuum emission. This maximum is in positional coincidence with the red lobe of the molecular outflow associated with core C1, suggesting that the HC$_{3}$N would trace not only the chemistry generated in the envelopes of the hot cores but also that related to the shocked gas \citep{hervias19}. The slightly elongated HC$_{3}$N feature extending towards the north-east from the bulk of the emission, which is in coincidence with the direction of the red lobe related to core C2, suggests the same interpretation.  

\subsubsection{H$_2$CS}

 Astrochemical modeling shows that the thioformaldehyde can be originated in the organosulfur chemistry that can be initiated in star-forming regions via the elementary gas-phase reaction of methylidyne radicals with hydrogen sulfide \citep{doddi20}. The H$_{2}$CS has been studied much less than its oxygen-substituted analog, the formaldehyde (H$_{2}$CO), however this molecular species has been used to study cores and outflows \citep{minh11,elakel22}. For instance, \citet{xu23} observed a H$_{2}$CS line with multicomponents which was used to estimate a temperature in a core embedded in a massive hub-filament system. In our case, the H$_{2}$CS is mainly concentrated in the core C1 with some surrounding extended emission (see Fig.\,\ref{molecFig} (d)). The core C4 presents weaker emission but well defined and correlated. It can be appreciated a faint structure towards the northwest in correspondence with the position of the outflow red-OC1 (see Fig.\,\ref{co-outflows}).  

\subsubsection{CH$_{3}$OH}
\label{metanol}

Methanol is a very important molecule that has been widely observed at both gas phase and solid state in the ISM (\citealt{qasim18} and references therein). As the authors indicated, it is generally accepted that CH$_{3}$OH formation is more efficient by solid state interactions on icy grain mantles, being the cold dense cores the suitable sites for its formation chemistry. It is known that in gas phase this species is precursor of several complex molecular species \citep{cecca17}.
Multiple lines in a wide range of frequencies, even in maser emission, are usually observed towards star-forming regions, and many of them used to trace molecular outflows (e.g. \citealt{bachiller95,palau07}). Shocks generated by jets and outflows are known to be efficient at boosting methanol to their gas phase \citep{rojas22}. Figure\,\ref{molecFig} (e) shows that all the analyzed cores exhibit CH$_{3}$OH emission. It is worth noting the methanol feature extending towards the north-east, which has a perfect morphological correspondence with the molecular outflow red-OC2 (see Fig.\,\ref{co-outflows}). This confirms the nature of this molecular feature studied with the $^{12}$CO emission in Sect.\,\ref{12CO} and suggests that the outflow activity is releasing CH$_{3}$OH from the solid state to the gas phase in the region. Additionally, some extended methanol emission appears towards the north-west, which may be related to the molecular outflow red-OC1.

\subsubsection{HNCO}
 
The isocyanic acid is a simple molecule containing the four main atoms essential for life as we know it, thus it can be considered as a prebiotic molecule. Indeed the smallest molecule possessing the biologically important amide bond, the formamide (NH$_{2}$CHO), seems to be in close relation with the chemistry of HNCO \citep{haupa19}. It was suggested that, while HNCO can be formed in the gas-phase during the cold stages of star formation, NH$_{2}$CHO forms most efficiently on the dust mantles, remaining frozen until the temperature rises enough to sublimate such icy mantles. The hydrogenation of HNCO is a likely formation route to lead NH$_{2}$CHO \citep{lopez-sepul15}. As Fig.\,\ref{molecFig} (f) shows, the HNCO emission is strongly concentrated at core C1. A small protrusion of weaker emission extends southwards containing cores C2 and C3, and also some weak emission seems to be associated with core C4. We suggest that the emission of this molecular species is tracing warm gas associated with the external layers of the cores.

\subsubsection{CN}
 
 The cyano radical (CN), one of the first detected interstellar molecular species \citep{mckellar40,adams41}, is a key molecule in many astrochemical chains. For instance, given that CN is very reactive with molecules possessing carbon double and triple bonds (C=C and C$\equiv$C respectively), it is involved in the formation of cyanopolyynes \citep{gans17} as the HC$_{3}$N presented in Sect.\,\ref{hc3n}. Figure\,\ref{molecFig} (g) displays the distribution of the CN emission showing that in general, CN maximums do not spatially coincide with the peaks of the continuum emission as found in other works \citep{beuther04,paron21}. In our case this can be appreciated mainly in cores C2 and C3.
 As \citet{beuther04} point out, this issue may be due to that the source embedded in the cores is at such an early
evolutionary stage that it does not generate enough UV photons to produce CN emission. Another possibility is that the lack of CN is due to its depletion related to the production of HC$_{3}$N. In any case, the CN emission would trace diffuse and extended gas surrounding the cores as it was found in several molecular cores by \citet{paron21}. However, given the extended features in the CN emission that seems
to coincide with the positions of the outflows, we made a detailed kinematic analysis of such an emission. Figure\,\ref{cn_outflows} shows the CN N=2--1, J=5/2--3/2 (F=5/2-3/2) emission distribution integrated between $-$80 and $-$70 \ks~(in blue) and between $-$55 and $-$40 \ks~(in red). The systemic velocity of the complex is about $-$64~\ks. 
It can be appreciated a similar morphology and kinematic as it was shown in the $^{12}$CO J=3--2 emission (see Fig.\,\ref{co-outflows}), suggesting that the CN is also tracing the molecular outflow activity related to the cores C1, C2, and C3. Moreover, the CN emission shows features likely related to red and blue lobes of a molecular outflow arising from the core C4. 

It is worth noting that 
the outflow cavity walls, which are narrow zones in between
the cold dense quiescent envelope material and the lower-density
warm cone where outflows are propagating, are pronounced in UV irradiation tracers such as the CN. Thus, the CN emission might highlight the border of such cavity walls \citep{tycho21}.
We conclude that we are presenting a very clear observational evidence that the CN traces the molecular gas related to the external part of the outflows, mainly to the cavities generated by them.

\begin{figure}[h]
   \centering
   \includegraphics[width=7.5cm]{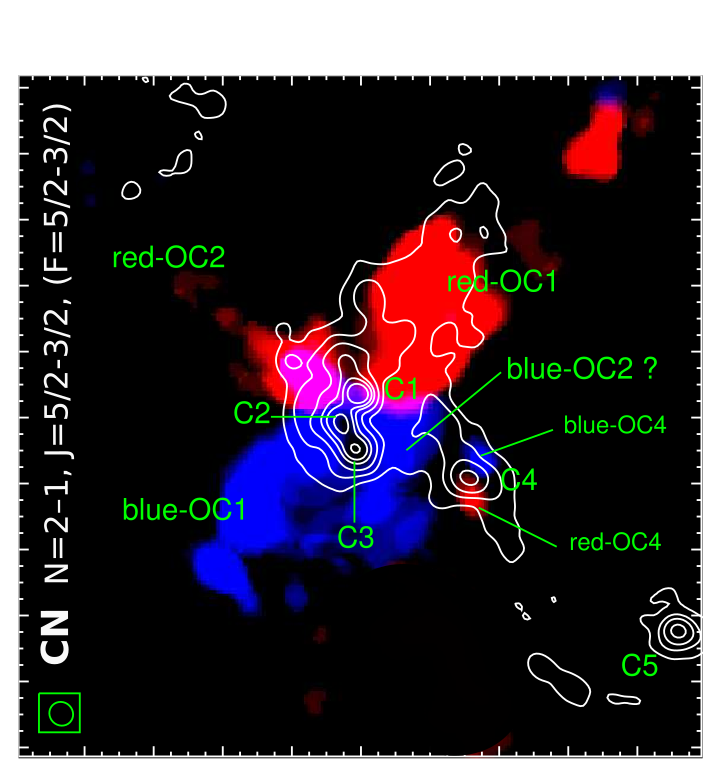}
    \caption{CN N=2--1, J=5/2--3/2 (F=5/2--3/2) emission distribution integrated between $-$80 and $-$70 \ks~(blue) and between $-$55 and $-$40 \ks~(red). The systemic velocity of the complex is about $-$64 \ks. The white contours represent the radio continuum at 340~GHz (12~m array). Levels are at 1, 10, 30, 60, 90, and 140~mJy beam$^{-1}$. The beam of the line emission is indicated at the bottom left corner.}
    \label{cn_outflows}
\end{figure}

\subsubsection{C$^{34}$S}
 
CS (carbon monosulfide) is the most ubiquitous among the sulfur-bearing molecules in the ISM. This molecular species was widely used to trace dense gas in star-forming regions 
(e.g. \citealt{bronf96}), and more recently in molecular filaments, hot and prestellar cores \citep{kimS20,zhou21,elakel22}. The less abundant isotopologue C$^{34}$S has been used to measure a possible $^{32}$S/$^{34}$S Galactic gradient \citep{yu20} and the CS depletion in prestellar cores \citep{kimS20}. The last authors found that the C$^{34}$S emission is not centrally peaked, or that the position where the intensity peaked is significantly shifted
when compared with the dust continuum maps, suggesting that the CS species became depleted significantly in the central high-density region of 
prestellar cores. In our case, we found that cores C1 and C4 traced in the dust continuum coincides with the C$^{34}$S peaks, while cores C2 and C3 lies in the region of the bulk of the extended emission (Fig.\,\ref{molecFig} (h)). Given that the investigated cores are active, we suggest that after the depletion of the CS molecules in the prestellar phase, the chemistry produced by the star-forming processes would contribute to increase the abundance of such sulfur-bearing molecular species. Additionally, the elongation in the C$^{34}$S emission towards the northeast and the feature extending towards the northwest, in coincidence with the molecular outflows red-OC2 and red-OC1, respectively, allow us to suggest that this molecule may be also a tracer of molecular outflows.

\subsubsection{HDO}
 
 Water is a fundamental requirement for life as we know it;  understanding its evolution, from its formation in molecular clouds to its presence in protoplanetary disks, is a challenge that aims to answer major questions; among them, if life can arise in other planetary systems.  
 However, H$_{2}$O emission lines are usually not observed from the ground, which often works with the rare isotopes: partially deuterated (HDO) and fully deuterated ($\rm D_{2}O$) water. Molecules tend to attach a D atom rather than an H atom because deuterated species have larger reduced masses and lower binding energies caused by the different zero-point vibrational energy \citep{vastel}, and this favors the production of species such as HD. The degree of deuterium fractionation in water is particularly related to the environmental conditions where it takes place \citep{jensen19} and serves 
 as a robust trail of the chemical and physical water evolution in star-forming regions \citep{ceccarelli}. The enrichment of species, such as HDO, is initiated by exothermic reactions, and therefore, the deuterium fractionation in water is expected to occur in the cold cloud molecular phase and later on the surface of dust grains \citep{kulczak,jensen21}. Figure\,\ref{molecFig} (i) displays the HDO emission map, centered and compacted only at core C1, for which high temperatures were derived (see Table\,\ref{ch3cn_DR_parameters}). Given its spatial distribution, we suggest that HDO emission comes from evaporated molecules due to the heating suffered by the ice layers on the grain surface. The release of this molecule into the gas-phase caused by desorption enriches again the environment in deuterated species as was studied in classical hot cores (\citealt{kulczak,csengeri19}). Based on the lack of HDO in the others cores, we suggest that in these regions such a molecular species would be  still frozen at the dust grains.

\section{Summary and Conclusion} 
\label{concl}

We present a study of the fragmentation and the star formation activity towards a massive molecular clump using high resolution and sensitivity ALMA data. The main goal of this work is to find evidence of high-mass star formation at core scale towards the massive clump AGAL G338.9188+0.5494 which harbours the EGO 338.92+0.55(b). 

The continuum emission at 340~GHz shows that the clump is fragmented into at least five cores, labeled from C1 to C5. The $^{12}$CO J=3--2 emission reveals the presence of molecular outflows arising from the cores C1, C2, and C4. C1 exhibits the more intense outflow activity. The molecular outflow related to core C1 is among the most massive (from 0.25 to 0.77~\msol) and energetic (from $0.4 \times 10^{46}$ to $1.2 \times 10^{46}$~erg) outflows, considering studies carried out with similar observations towards this type of sources.

Interestingly, the cyanide radical  extended emission exhibits the same morphology and kinematics than the $^{12}$CO J=3--2 emission, suggesting that the CN molecule is also tracing the same molecular outflow activity.  Given that the CN is an UV irradiation tracer, we point out that its emission highlights the border of the cavity walls carved out by the outflows.

The CH$_3$CN J=13--12 (K=2) moment 1 map shows a clear velocity gradient towards the core C1, attributable to a rotating disk, whose direction is perpendicular to the molecular outflow direction. 

The rotational diagrams for CH$_3$CN and CH$_3$CCH, yields temperatures of about 340 and 72~K, respectively, for the core C1. This suggests that the methyl cyanide would be placed closer to the protostar than the methyl acetylene, which would be tracing outermost layers of gas. 
Using a range of temperatures going from 120 K (about the typical molecular desorption temperature in hot cores) to  340~K obtained from the rotational diagram of the CH$_3$CN, the mass of core C1 ranges from 3 to 10 \msol.
We point out that the use of typical desorption temperatures or temperatures derived from  molecular species such as methyl cyanide, tracing the gas at core scales, is more appropriated to characterize cores than using the typical dust temperatures obtained from the clump scales.

The mid-IR 4.5~$\mu$m and near-IR K${\rm _s}$ band extended emissions coincide in position and inclination with the molecular outflow of core C1, in particular with the blue-shifted lobe. We suggest that the molecular outflow activity related to core C1 is the main responsible for the brightness of the EGO 338.92+0.55(b) at 4.5~$\mu$m. Therefore, the counterpart of the EGO, at core scale, should be a molecular outflow with average mass and energy of about 0.5~\msol~and $10^{46}$~erg, respectively. 

Based on the estimated accretion rate and the dynamical age of the outflow towards core C1, we suggest that the protostar forming inside the core would have at most 0.4~\msol. Therefore, considering that the mass of the core is at most 10~\msol, and that we did not find any evidence of accreting gas filaments, we conclude that it is unlikely  a high-mass star forms within this core.

 \begin{acknowledgements}

We thank the anonymous referee for her/his useful comments and corrections. M.O. and S.P. are members of the Carrera del Investigador Cient\'\i fico of CONICET, Argentina.  N.I. is posdoctoral fellow and N.M. and A.M. are  doctoral fellows of CONICET, Argentina.
This work was partially supported by the Argentina grant PIP 2021 11220200100012 from CONICET.
This work is based on the following ALMA data: ADS/JAO.ALMA $\#$ 2015.1.01312, and 2017.1.00914. ALMA is a partnership of ESO (representing its member states), NSF (USA) and NINS (Japan), together with NRC (Canada), MOST and ASIAA (Taiwan), and KASI (Republic of Korea), in cooperation with the Republic of Chile. The Joint ALMA Observatory is operated by ESO, AUI/NRAO and NAOJ.

\end{acknowledgements}

%
%

\bibliographystyle{aa}  
\bibliography{ref}
\IfFileExists{\jobname.bbl}{}
{\typeout{}
\typeout{****************************************************}
\typeout{****************************************************}
\typeout{** Please run "bibtex \jobname" to optain}
\typeout{** the bibliography and then re-run LaTeX}
\typeout{** twice to fix the references!}
\typeout{****************************************************}
\typeout{****************************************************}
\typeout{}
}
\label{lastpage}

\begin{appendix}

\section{Band 6 spectra towards core C1}
\label{appendix}

Figures\,\ref{band6spw0}, \ref{band6spw1}, \ref{band6spw2}, and \ref{band6spw3} display the spectra of band 6 spectral windows 0, 1, 2, and 3, respectively extracted from a beam centered at the position of core C1. A tentative molecular line identification was done using CASA software cross checking with the JPL and CDMS databases using the Splatalogue.

 \begin{figure*}[h]
   \centering
   
   \includegraphics[width=19cm]{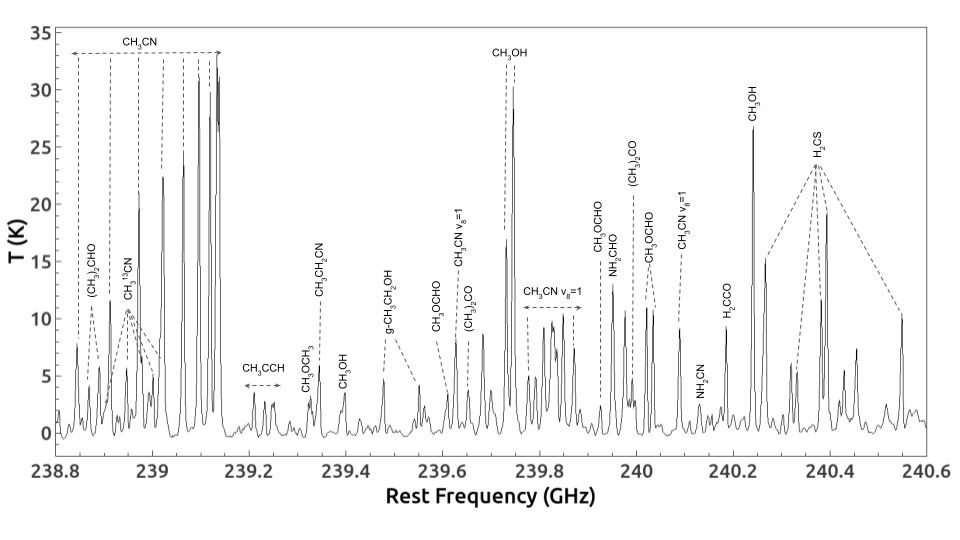}
    \caption{Band 6 spw0 towards core 1.}
    \label{band6spw0}
\end{figure*}

 \begin{figure*}[h]
   \centering
   
   \includegraphics[width=19cm]{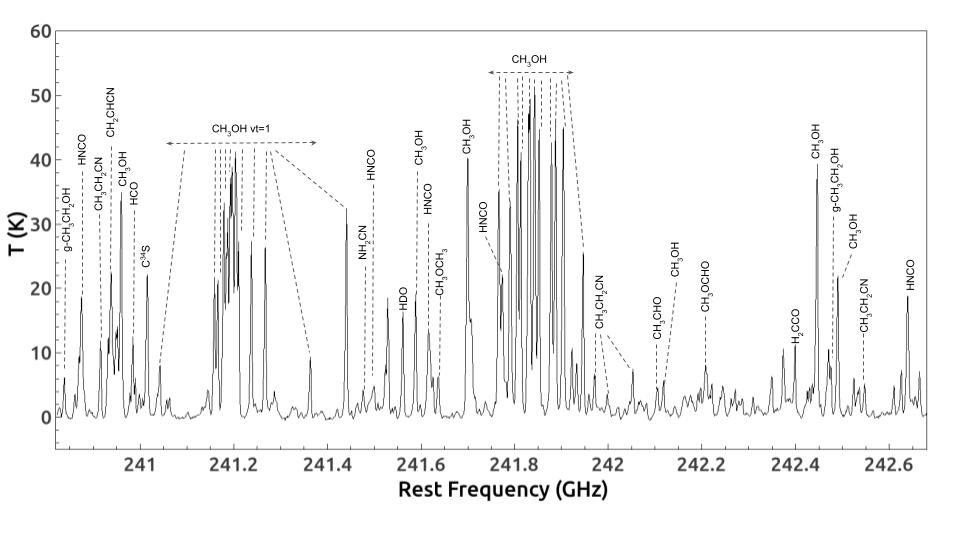}
    \caption{Band 6 spw1 towards core 1.}
    \label{band6spw1}
\end{figure*}

 \begin{figure*}[h]
   \centering
   
   \includegraphics[width=19cm]{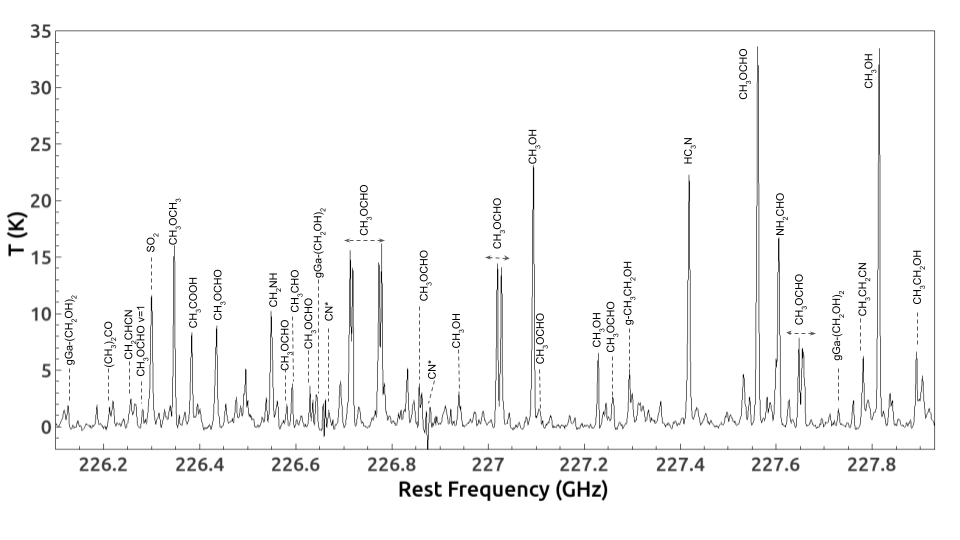}
    \caption{Band 6 spw2 towards core 1. (*) The CN emission is very weak towards this core.}
    \label{band6spw2}
\end{figure*}

 \begin{figure*}[h]
   \centering
   
   \includegraphics[width=19cm]{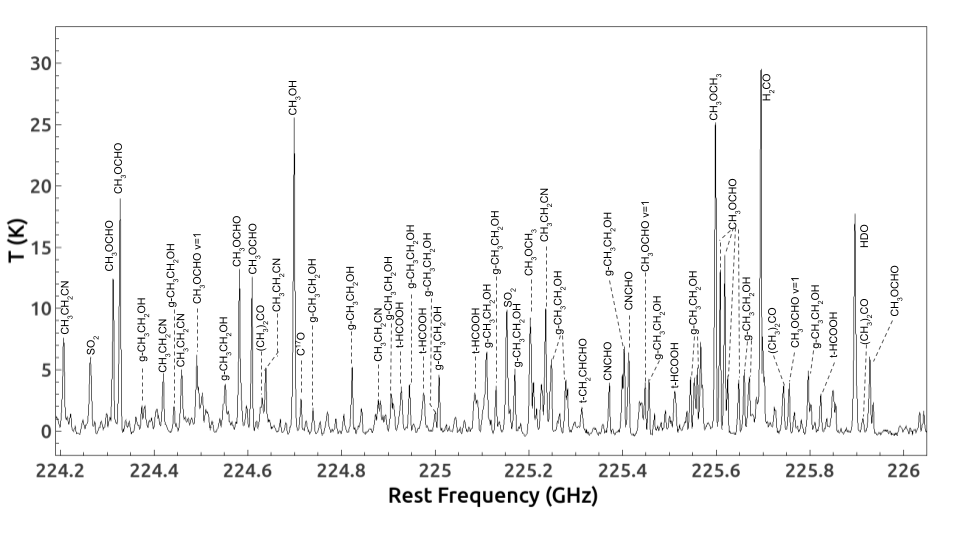}
    \caption{Band 6 spw3 towards core 1.}
    \label{band6spw3}
\end{figure*}

\end{appendix}

\end{document}